\documentclass[prd,twocolumn,nofootinbib,aps,tightenlines,preprintnumbers,notitlepage,longbibliography,superscriptaddress]{revtex4-1}
\usepackage{amsmath, amssymb}
\usepackage{cellspace}
\usepackage{color}
\usepackage{float}
\usepackage{graphicx}
\usepackage{natbib}
\usepackage[colorlinks=true,citecolor=blue,urlcolor=blue]{hyperref}

\newcommand{\vp}{v_\textrm{ph}}
\newcommand{\fhg}{{}_2 F _1}

\begin{document}

\preprint{MIT-CTP/5222}

\title{Subluminal stochastic gravitational waves in pulsar-timing arrays and astrometry}

\author{Wenzer Qin}
\affiliation{Center for Theoretical Physics, Massachusetts Institute of Technology, Cambridge, Massachusetts 02139, USA}
\affiliation{Department of Physics \& Astronomy, Johns Hopkins University, Baltimore, MD 21218, USA}
\author{Kimberly K. Boddy}
\affiliation{Theory Group, Department of Physics, The University of Texas at Austin, Austin, TX 78712, USA}
\author{Marc Kamionkowski}
\affiliation{Department of Physics \& Astronomy, Johns Hopkins University, Baltimore, MD 21218, USA}

\begin{abstract}
The detection of a stochastic background of low-frequency gravitational waves by pulsar-timing and astrometric surveys will enable tests of gravitational theories beyond general relativity.
These theories generally permit gravitational waves with non-Einsteinian polarization modes, which may propagate slower than the speed of light.
We use the total-angular-momentum wave formalism to derive the angular correlation patterns of observables relevant for pulsar timing arrays and astrometry that arise from a background of subluminal gravitational waves with scalar, vector, or tensor polarizations.
We find that the pulsar timing observables for the scalar longitudinal mode, which diverge with source distance in the luminal limit, are finite in the subluminal case.
Furthermore, we apply our results to $f(R)$ gravity, which contains a massive scalar degree of freedom in addition to the standard transverse-traceless modes.
The scalar mode in this $f(R)$ theory is a linear combination of the scalar-longitudinal and scalar-transverse modes, exciting only the monopole and dipole for pulsar timing arrays and only the dipole for astrometric surveys.
\end{abstract}

\maketitle

%%%%%%%%%%%%%%%%%%%%%%%%%%%%%%%%%%%%%%%%%%%%%%%%%%%%%%%%%%%%%%%%%%%%%%%%%%%%%%%
\section{Introduction}

There are world-wide efforts to detect a stochastic background of gravitational waves (GWs) using pulsar timing arrays (PTAs)~\cite{Verbiest:2016vem,Perera:2019sca,Hobbs:2013aka,Manchester:2012za,Lentati:2015qwp}.
Due to the rotation of the pulsars, beams of radiation from the stars may periodically sweep through the Earth's line of sight and appear as regular pulses of light.
The presence of GWs then modifies the expected pulse arrival times at Earth.
For a background of GWs, the pulse arrival times from different pulsars are correlated across the sky; in particular, a stochastic background produces an angular correlation given by the the Hellings-Downs curve~\cite{Hellings:1983fr}.%
\footnote{The NANOGrav Collaboration has found strong evidence of a stochastic process in pulsars; however, they find no significant evidence of quadrupolar correlations which would be characteristic of a background of GWs~\cite{Arzoumanian:2020vkk}.}
GWs can also induce a shift in the apparent position of stars, which may be observed in astrometric surveys~\cite{Braginsky:1989pv,Kaiser:1996wk}.
Similar to PTAs, the stellar shift exhibits particular angular correlations from a stochastic GW background~\cite{Book:2010pf}.

The GWs in general relativity (GR) arise from the transverse-traceless tensor modes of the metric perturbation.
If, however, GR is modified, there may be additional propagating degrees of freedom from the scalar and vector modes, leading to GW polarization states beyond the standard two.
As a result, the angular correlation of pulse arrival times for PTAs or stellar positions for astrometry has a different functional form that depends on the GW polarization and the relative amplitudes of the polarization states.
There are generically six polarizations states, which we classify as follows: 2 transverse-traceless tensor modes, 2 vector modes, a scalar-longitudinal mode (SL), and a scalar-transverse mode (ST).
Previous studies have calculated the normalized, individual contributions to the angular correlation function due to all six polarizations for PTA~\cite{Lee:2008aa,Chamberlin:2011ev,Gair:2014rwa,Gair:2015hra,Qin:2018yhy} and astrometry observables~\cite{Mihaylov:2018uqm,OBeirne:2018slh,Qin:2018yhy,Mihaylov:2019lft}.
The relative amount each mode contributes depends on the particular theory.
Constraints on alternative theories of gravity using PTAs has also been considered in Refs.~\cite{Cornish:2017oic} and \cite{OBeirne:2019lwp}.
While most of these previous calculations assume all GW modes propagate at the speed of light, there are scenarios in which certain modes may experience subluminal propagation. In particular, the prospect of detecting massive gravitons with PTAs has been considered by Refs.~\cite{Baskaran:2008za,Lee:2010cg,Lee:2014awa}.

In this paper, we derive the auto- and cross-correlation patterns for PTA and astrometry observables due to subluminal GW polarization modes under the total-angular-momentum (TAM) formalism~\cite{Dai:2012bc}, following the methods outlined in Ref.~\cite{Qin:2018yhy}.
We then investigate a particular form of $f(R)$ gravity that contains a single additional degree of freedom in the form of a massive scalar field.
The speed of propagation then depends on the mass of the field, and the associated GW polarization is comprised of a superposition of the SL and ST modes.

Although observations from LIGO constrain the propagation speed of the transverse-traceless modes to between $1-3 \times 10^{-15}$ and $1+7 \times 10^{-16}$ times the speed of light~\cite{Monitor:2017mdv}, it may be possible for an alternative theory of gravity to possess a frequency-dependent propagation speed.
In particular, LIGO studies GWs of frequency $\sim 10^2~\mathrm{Hz}$, while LISA will probe frequencies of $\sim 10^{-2}~\mathrm{Hz}$, and PTAs and astrometry probe frequencies of $\sim \mathrm{nHz}$; thus, it is possible that even if LIGO does not detect subluminal GWs, other observatories could.
Therefore, we consider subluminal propagation of transverse-traceless modes for completeness.
The same argument holds true for scalar and vector modes.
To date, LIGO has found no evidence for non-Einsteinian polarizations~\cite{Abbott:2017oio,Isi:2017fbj,Abbott:2018utx}; however, even if future studies were able to constrain the velocities of these modes, GWs at frequencies below LIGO's range of sensitivity could avoid these bounds.

We note that recent work has derived the astrometric angular correlation functions and power spectra for nonluminal GW propagation~\cite{Mihaylov:2019lft}.
Our results for subluminal GWs numerically agree with and complement those in Ref.~\cite{Mihaylov:2019lft}, which were derived using different methods described in Ref.~\cite{Mihaylov:2018uqm}.

The outline of this paper is as follows.
In Sec.~\ref{sec:TAM}, we give a brief introduction to the TAM wave formalism.
In Sec.~\ref{sec:correlation}, we calculate the general expressions for the PTA and astrometric angular response to a subluminal GW with nonstandard polarizations.
We then turn to $f(R)$ gravity as a concrete example in Sec.~\ref{sec:fR} and relate the scalar degree of freedom to a particular combination of ST and SL modes.
We conclude in Sec.~\ref{sec:conclusions}.

%%%%%%%%%%%%%%%%%%%%%%%%%%%%%%%%%%%%%%%%%%%%%%%%%%%%%%%%%%%%%%%%%%%%%%%%%%%%%%%
\section{Total-angular-momentum waves}
\label{sec:TAM}

In many studies of cosmological perturbations and stochastic gravitational wave backgrounds, the spacetime metric perturbation is decomposed into plane waves, as these provide a simple and familiar orthonormal basis.
However, the simplicity is lost once these waves are projected onto spherical surfaces such as the sky.
A more natural basis in these situations is to work with eigenstates of total angular momentum, i.e. TAM waves~\cite{Dai:2012bc}.

For example, scalar fields can be decomposed in terms of plane waves $e^{i \boldsymbol{k} \cdot \boldsymbol{x}}$ or scalar TAM waves
\begin{equation}
  \Psi_{\ell m}^k (\boldsymbol{x}) = j_\ell (kr) Y_{\ell m} (\boldsymbol{\hat{n}}) ,
\end{equation}
where the $j_\ell (kr)$ are spherical Bessel functions, $Y_{\ell m} (\boldsymbol{\hat{n}})$ are scalar spherical harmonics, and $\boldsymbol{k}$ is the wave number of the GW with amplitude $k$.
We have defined $\boldsymbol{x} = r \boldsymbol{\hat{n}}$, where $r$ is the distance from Earth in the direction $\boldsymbol{\hat{n}}$ of the pulsar.
Similarly, tensors fields can be written in terms of plane waves $\varepsilon^s_{ab} (\boldsymbol{k}) e^{i \boldsymbol{k} \cdot \boldsymbol{x}}$, where $\varepsilon^s_{ab} (\boldsymbol{k})$ is the polarization tensor for plane waves of polarization $s$, or TAM waves $\Psi^{k, \alpha}_{(\ell m) ab} (\boldsymbol{x})$, which are combinations of spherical Bessel functions and tensor spherical harmonics $Y^{\alpha}_{(\ell m)ab} (\boldsymbol{\hat{k}})$ of polarization $\alpha$.
The subscripts $a$ and $b$ are abstract spatial indices.
The exact expressions for the TAM waves can found in Eq.~(94) of Ref.~\cite{Dai:2012bc}.

We can convert between the tensor plane wave and TAM wave bases using
\begin{equation}
  \varepsilon^s_{ab} (\boldsymbol{k}) e^{i \boldsymbol{k} \cdot \boldsymbol{x}}
  = 4 \pi \sum_{\alpha, \ell, m} i^\ell B^\alpha_{(\ell m)} (\hat{k}) \Psi^{\alpha,k}_{(\ell m) ab} (\boldsymbol{x}) ,
  \label{eqn:plane_to_TAM_tensor}
\end{equation}
where the coefficients $B^\alpha_{(\ell m)}$ are given by
\begin{equation}
  B^\alpha_{(\ell m)} = \varepsilon_s^{ab} (\boldsymbol{k}) \left[Y^{\alpha}_{(\ell m)ab} (\boldsymbol{\hat{k}})\right]^\ast .
\end{equation}

%%%%%%%%%%%%%%%%%%%%%%%%%%%%%%%%%%%%%%%%%%%%%%%%%%%%%%%%%%%%%%%%%%%%%%%%%%%%%%%
\section{Power spectra}
\label{sec:correlation}

The geodesic of observed light emanating from a source may be altered by a GW passing between the source and Earth.
A stochastic background of GWs is expected to induce particular angular correlation patterns for PTA and astrometry observables.
In PTAs, the GW affects the light travel time from a pulsar and thus affects the observed time of arrival.
We choose, however, to work with the relative shift in the pulse arrival frequency $z(t,\boldsymbol{\hat{n}})$, rather than the arrival time, to make connections to previous literature.
We note that this change in the observable impacts the time domain information, but not the angular response of the signal~\cite{Qin:2018yhy}.
We may express the shift as an expansion in spherical harmonics,
\begin{equation}
  z(t,\boldsymbol{\hat{n}})
  = \sum_{\ell, m} z_{\ell m}(t) Y_{\ell m}(\boldsymbol{\hat{n}}) ,
\end{equation}
for a pulsar located in the $\boldsymbol{\hat{n}}$ direction at time $t$.
In astrometry, the GW affects the apparent location of stars.
The resulting shift in position may be expanded as
\begin{equation}
  \delta^a (t,\boldsymbol{\hat{n}}) = \sum_{\ell, m}
  \left[ E_{\ell m}(t) Y_{\ell m}^{E,a} (\boldsymbol{\hat{n}})
    + B_{\ell m}(t) Y_{\ell m}^{B,a} (\boldsymbol{\hat{n}}) \right] ,
\end{equation}
where $Y_{\ell m}^{E,a}$ and $Y_{\ell m}^{B,a}$ are vector spherical harmonics~\cite{Dai:2012bc}.

The correlation functions and power spectra for these observables are derived in Ref.~\cite{Qin:2018yhy} using the TAM formalism~\cite{Dai:2012bc}, under the assumption that all polarization modes of the GWs propagate at the speed of light with the GW frequency equaling its wave number, $\omega = k$ (with $c=1$).
Here, we consider the more general case of a GW of polarization $\alpha$ with a dispersion relation $\omega_\alpha(k)$.
For massive gravity models, where the propagating mode behaves like a particle of mass $m_\alpha$, this dispersion relation is given by $\omega_\alpha^2(k) = k^2 + m_\alpha^2$, which we assume for the remainder of the paper.
We define the phase velocity $v_{\textrm{ph},\alpha} \equiv \omega_\alpha / k$ and the group velocity $v_\alpha \equiv d\omega_\alpha / dk = k/\omega_\alpha$; thus, while the group velocity is subluminal, the phase velocity is superluminal.

We expand the metric perturbation as
\begin{equation}
  h_{ab}(t,\boldsymbol{x}) =
  \int \frac{k^2\, dk}{(2\pi)^3} 4\pi i^\ell h_{\ell m}^\alpha(k)
  \Psi_{(\ell m)ab}^{\alpha,k} (\boldsymbol{x}) e^{-i\omega_\alpha(k) t} ,
  \label{eq:metric-pert}
\end{equation}
for a single TAM wave $\Psi_{(\ell m)ab}^{\alpha,k}$ with amplitude $h_{\ell m}^\alpha$.
Using this expansion, we write the power spectra from a stochastic GW background as~\cite{Qin:2018yhy}
\begin{equation}
  C_\ell^{XX',\alpha} \propto 32\pi^2
  F_\ell^{X,\alpha} \left(F_\ell^{X',\alpha}\right)^*,
  \label{eq:Cl}
\end{equation}
corresponding to the PTA and astrometry observables $X,X' \in \{z,E,B\}$.
In this expression, we have omitted a factor that encompasses time domain information, including the dependence on the GW frequency and the cadence of the observation.
As we show in the following subsections, the projection factors $F_\ell^{X,\alpha}$ (i.e.\ detector response functions) depend on the phase velocity and thus cannot be factored out from the integral over $k$ in Eq.~\eqref{eq:metric-pert}, unlike the case for luminal GWs.
Therefore, the actual power spectrum receives contributions from a range of velocities, determined by the window function used for observation.
For the purposes of this work, we assume the window function is narrow so that Eq.~\eqref{eq:Cl} holds; our results may be applied to the full expression of Eq.~\eqref{eq:Cl} in Ref.~\cite{Qin:2018yhy} for more general cases.
In the following subsections, we derive the expressions for the projection factors, in close parallel with Ref.~\cite{Qin:2018yhy}.

%%%%%%%%%%%%%%%%%%%%%%%%%%%%%%%%%%%%%%%%
\subsection{Pulsar Timing Arrays}

The fractional shift in the observed pulse frequency of a pulsar due to a metric perturbation $h_{ab}$ is
\begin{equation}
  z(t,\boldsymbol{\hat{n}}) = -\frac{1}{2} n^a n^b \int_{t}^{t-r_s} dt'\
  h_{ab,0}[t',(t-t')\boldsymbol{\hat{n}}],
\end{equation}
where $\boldsymbol{\hat{n}}$ is the direction of the pulsar in the sky, $t$ is the observation time of a pulse, and $r_s$ is the distance to the pulsar.
The time derivative acts only on the explicit time dependence in Eq.~\eqref{eq:metric-pert}.
Additionally, in the TAM formalism,
\begin{equation}
  n^a n^b \Psi_{(\ell m) ab}^{\alpha,k} (\boldsymbol{x})
  = -R_\ell^{L,\alpha}(kr) Y_{\ell m}(\boldsymbol{\hat{n}}),
\end{equation}
where $Y_{\ell m}$ are spherical harmonics and $R_\ell^{L,\alpha}$ are radial functions, given in Appendix~\ref{sec:derive}.
Thus, the shift in pulse frequency becomes
\begin{equation}
  z(t,\boldsymbol{\hat{n}}) =
  4\pi i^\ell \int \frac{k^2\, dk}{(2\pi)^3} h_{\ell m}^\alpha(k) F_\ell^{z,\alpha}
  Y_{\ell m}(\boldsymbol{\hat{n}}) e^{-i\omega_\alpha(k) t} .
\end{equation}
The projection factor $F_\ell^{z,\alpha}$ is analogous to that in Ref.~\cite{Qin:2018yhy} for luminal GWs and is given by
\begin{equation}
  F_\ell^{z,\alpha} \equiv -\frac{iv_{\textrm{ph},\alpha}}{2}
  \int_0^\infty dx\ R_\ell^{L,\alpha}(x) e^{ixv_{\textrm{ph},\alpha}} ,
  \label{eqn:Fz}
\end{equation}
where we have taken the distant-source limit, $kr_s\to\infty$.

%%%%%%%%%%%%%%%%%%%%%%%%%%%%%%%%%%%%%%%%
\subsection{Astrometry}

The astrometric deflection due to a metric perturbation $h_{ab}$ is
\begin{widetext}
\begin{equation}
  \delta^a (\boldsymbol{\hat{n}},t) = \Pi^{ac} n^b \left\{-\frac{1}{2}
  h_{bc}(t,\boldsymbol{0})+ \frac{1}{r_s} \int_0^{r_s} \, dr\, \left[
    h_{bc}(t-r,r\boldsymbol{\hat{n}})  - \frac{r_s-r}{2} n^d\partial_c
    h_{bd}(t-r,r\boldsymbol{\hat{n}}) \right] \right\},
\end{equation}
where $\Pi_{ab}(\boldsymbol{\hat{n}})=\eta_{ab}-\hat{n}_a \hat{n}_b$ projects onto the plane orthogonal to $\boldsymbol{\hat{n}}$.
Expanding the metric perturbation in terms of TAM waves, we obtain
\begin{equation}
  \delta^a (\boldsymbol{\hat{n}},t) = \sum_{\ell,m} \sum_{\alpha} 4\pi i^\ell
  \int \frac{k^2\, dk}{(2\pi)^3} h_{\ell m}^\alpha (k)
  \left[ F_\ell^{E,\alpha} Y_{\ell m}^{E,a}(\boldsymbol{\hat{n}})
    + F_\ell^{B,\alpha} Y_{\ell m}^{B,a}(\boldsymbol{\hat{n}}) \right]
  e^{-i\omega_\alpha(k) t} ,
\end{equation}
where
\begin{align}
  F_\ell^{E,\alpha} &= -\frac{1}{2} R_\ell^{E,\alpha}(0) + \int_0^\infty dx\
  \left[R_\ell^{E,\alpha}(x) - \frac{1}{2}\sqrt{\ell (\ell + 1)} R_\ell^{L,\alpha} \right] \frac{1}{x} e^{ixv_{\textrm{ph},\alpha}} \label{eqn:FE} \\
  F_\ell^{B,\alpha} &= \int_0^\infty dx\
  R_\ell^{B,\alpha}(x) \frac{1}{x} e^{ixv_{\textrm{ph},\alpha}}
  \label{eqn:FB}
\end{align}
\end{widetext}
in the distant-source limit.

%%%%%%%%%%%%%%%%%%%%%%%%%%%%%%%%%%%%%%%%
\subsection{Power spectra}
\label{sec:power_spectra}

\begin{figure*}[htbp]
  \centering
  \includegraphics[width=0.95\linewidth]{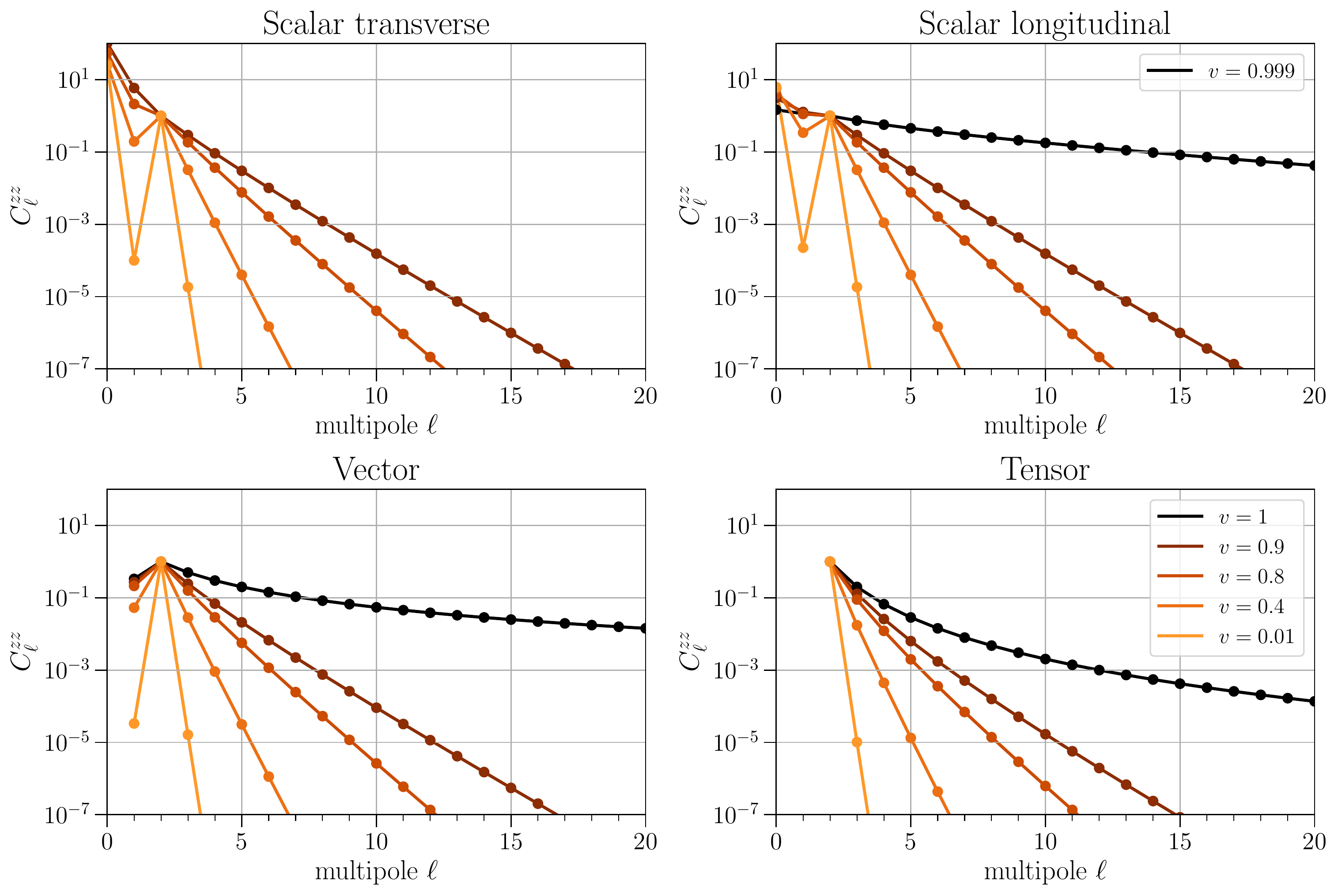}
  \caption{The $C^{zz}_\ell$ power spectra for the scalar, vector, and tensor GW polarization modes at various values of the group velocity $v$, as indicated in the legend of the lower-right panel.
    The spectra are normalized to $C^{zz}_2$.
    The $v=1$ line for the ST mode is not shown, since it has contributions from $\ell=0$ and $1$ only.
    The SL mode is divergent for $v=1$ due to photons surfing the GW wave, so we show $v=0.999$ instead.}
  \label{fig:czz}
\end{figure*}

\begin{figure*}[htbp]
  \centering
  \includegraphics[width=0.95\linewidth]{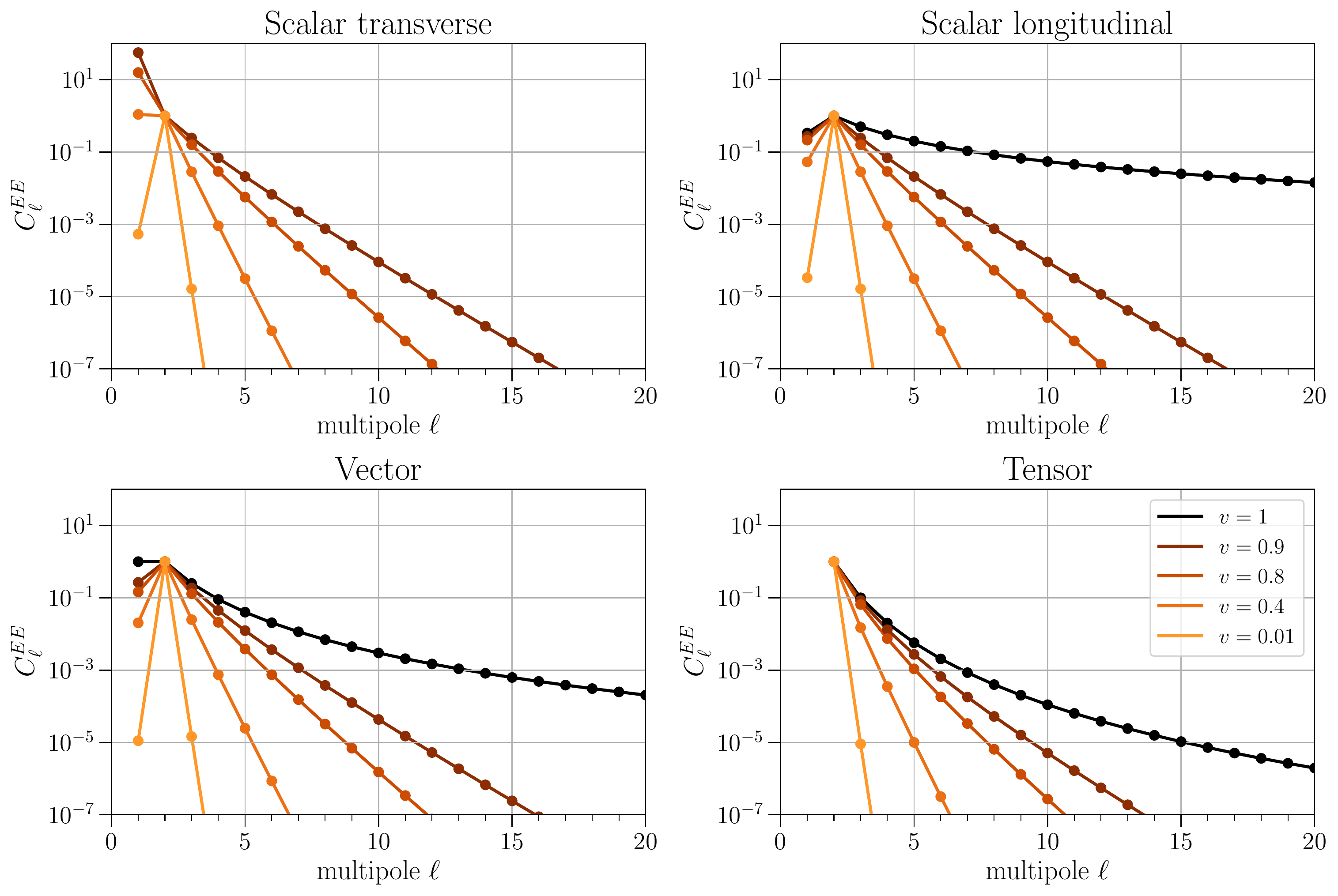}
  \caption{The $C^{EE}_\ell$ power spectra for the scalar, vector, and tensor GW polarization modes at various values of the group velocity $v$, as indicated in the legend of the lower-right panel.
    The spectra are normalized to $C^{EE}_2$.
    The $v=1$ line for the ST mode is not shown, since it has contributions from $\ell=1$ only.}
  \label{fig:cee}
\end{figure*}

\begin{figure*}[htbp]
  \centering
  \includegraphics[width=0.95\linewidth]{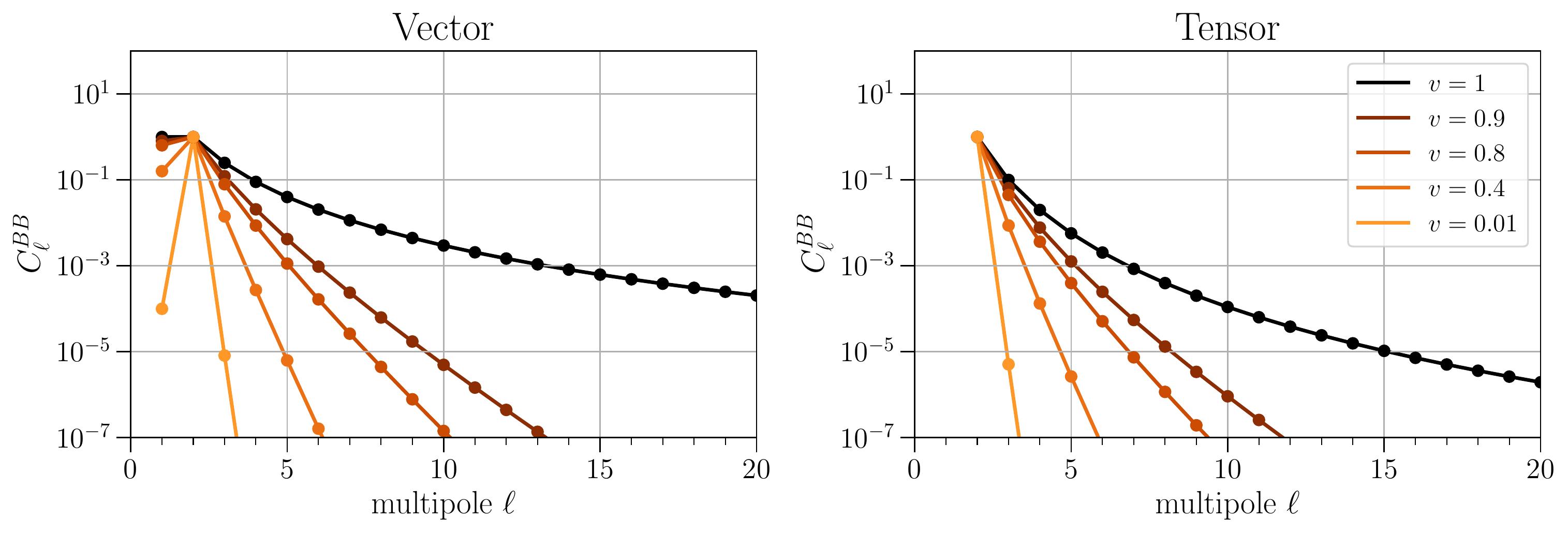}
  \caption{The $C^{BB}_\ell$ power spectra for the vector and tensor GW polarization modes at various values of the group velocity $v$, as indicated in the legend of the right panel.
    The spectra are normalized to $C^{BB}_2$.
    The scalar polarizations do not generate $B$-mode deflections.}
  \label{fig:cbb}
\end{figure*}

\begin{figure*}[htbp]
  \centering
  \includegraphics[width=0.95\linewidth]{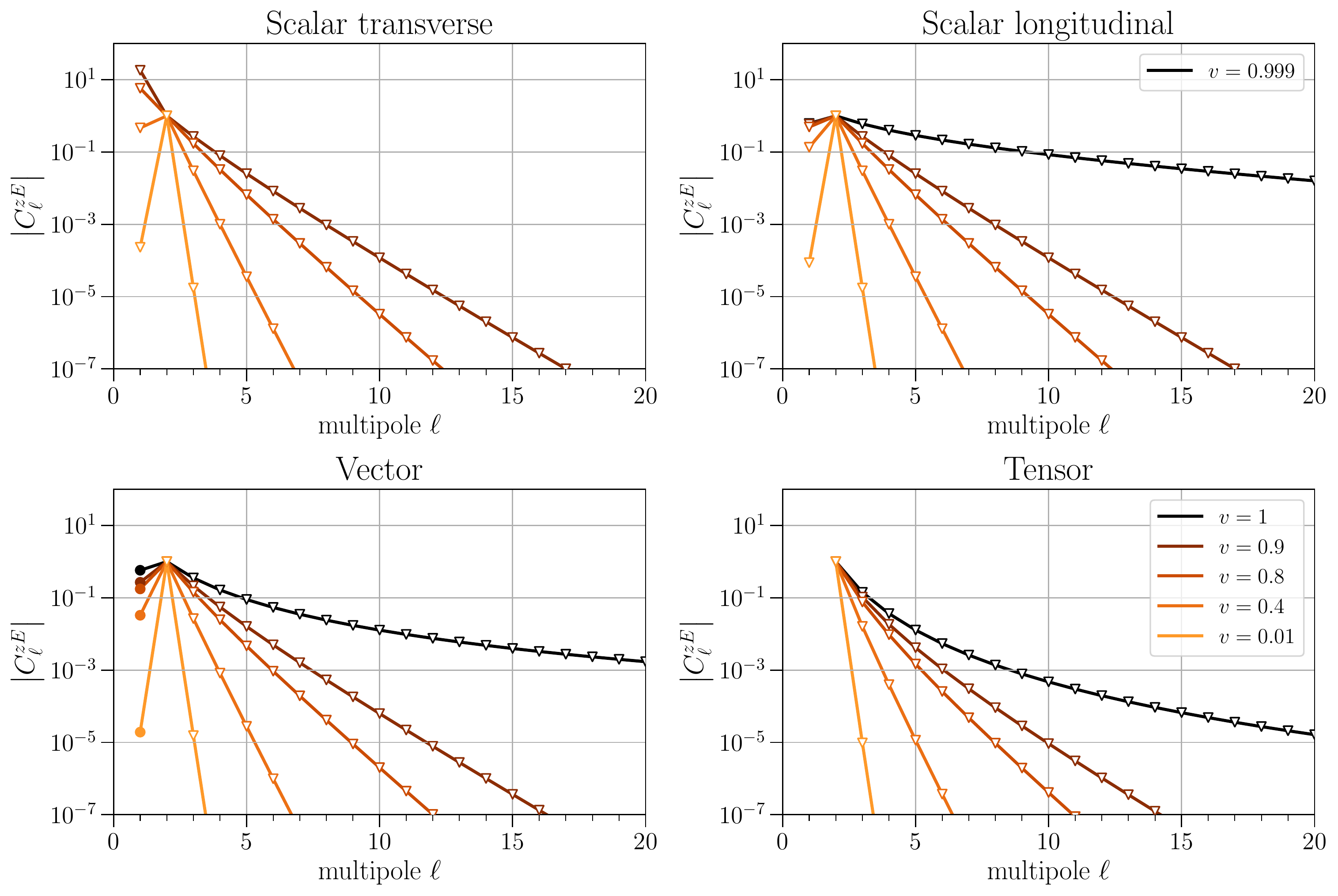}
  \caption{The absolute value of the $C^{zE}_\ell$ power spectra for the scalar, vector, and tensor GW polarization modes at various values of the group velocity $v$, as indicated in the legend of the lower right panel.
    The spectra are normalized to $\left|C^{zE}_2\right|$.
    The $\triangledown$ and $\bullet$ points indicate the power spectrum is negative and positive, respectively.
    The $v=1$ line for the ST mode is not shown, since it has contributions from $\ell=1$ only.
    The SL mode is divergent for $v=1$ due to photons surfing the GW wave, so we show $v=0.999$ instead.}
  \label{fig:cze}
\end{figure*}

We derive the analytic expressions for the projection factors $F_\ell^{z,\alpha}$, $F_\ell^{E,\alpha}$, and $F_\ell^{B,\alpha}$ in Appendix~\ref{sec:derive} and summarize the results in Table~\ref{tab:F}.
We show the resulting power spectra $C^{zz}_\ell$, $C^{EE}_\ell$, $C^{BB}_\ell$, and $C^{zE}_\ell$ as functions of the multipole $\ell$ for each GW polarization in Figs.~\ref{fig:czz}, \ref{fig:cee}, \ref{fig:cbb}, and \ref{fig:cze}, respectively.
For each case, we compare the power spectra at group velocities $v \in \{0.01, 0.4, 0.8, 0.9\}$ to the power spectra at $v=1$ from Ref.~\cite{Qin:2018yhy}.
The $C^{zz}_\ell$ and $C^{zE}_\ell$ spectra, however, diverge for SL modes in the $kr\to\infty$ limit for $v=1$, due to the light ray surfing the GW; therefore, we show $v=0.999$ rather than $v=1$, since there is no surfing for subluminal GWs and the projection factor $F^{z,\textrm{SL}}_\ell$ is finite.
Note that we have dropped the polarization subscript $\alpha$ on $v_\alpha$ for notational simplicity, with the understanding that each GW polarization mode may propagate with its own distinct frequency.

We normalize all power spectra by their quadrupole contribution.
For $v=1$, the power spectra either have just one or two dominant contributions at low multipoles, while higher multipoles are either suppressed by factors of at least $\sim \ell^2$ or vanish altogether for the ST mode (see Table~1 of Ref.~\cite{Qin:2018yhy}).%
\footnote{The ST mode has a monopole and dipole power spectrum in the formal limit of $kr_s\to\infty$. In reality, the finite distance to the photon source produces nonzero power at higher multipoles, but we expect the effect to be small and ignore it for the purposes of our discussion.}
As the ST mode has no quadrupole to set the normalization, we omit the $v=1$ curve (formed by the monopole and dipole) from Fig.~\ref{fig:czz} and the single $v=1$ point (from the dipole) from Figs.~\ref{fig:cee} and \ref{fig:cze}.

For subluminal propagation, all power spectra become increasingly dominated by the quadrupole (and the monopole, for the scalar modes of $C^{zz}_\ell$) as $v$ decreases.
We investigate this behavior in the $v \rightarrow 0$ limit in Appendix~\ref{sec:derive}, but we may also understand it using the plane-wave basis for GWs.
The pulsar frequency shift and angular deflection for luminal GWs are
\begin{align}
  z(\boldsymbol{\hat{n}}) &= \frac{n^a n^b h_{ab}}{2 (1 + \boldsymbol{\hat{k}} \cdot \boldsymbol{\hat{n}} )} , \\
  \delta^a (\boldsymbol{\hat{n}}) &=  \frac{(n^a + k^a) n^b n^c h_{bc}}{2 (1 + \boldsymbol{\hat{k}} \cdot \boldsymbol{\hat{n}})} - \frac{1}{2} n^b h_{ab} ,
\end{align}
where $\boldsymbol{\hat{k}}$ denotes the direction of GW propagation~\cite{Book:2010pf}.
The factors of $1 + \boldsymbol{\hat{k}} \cdot \boldsymbol{\hat{n}}$ are derived by assuming the wave propagates at the speed of light; if we instead have waves propagating with $v<1$, these factors become $1 + v \boldsymbol{\hat{k}} \cdot \boldsymbol{\hat{n}}$, which go to unity as $v \rightarrow 0$.
Then $z(\boldsymbol{\hat{n}})$ and $ \delta^a (\boldsymbol{\hat{n}})$ reduce to projections of the GW polarizations onto the sky, which for the tensor and vector modes are quadrupolar in form, and for the scalar modes have both monopole and quadrupole components.

Notably, this trend holds true for the ST mode, for which power at $\ell > 1$ no longer vanishes.
There is a stark contrast between the power spectra $C^{zz}_\ell$ of the ST mode with $v=1$ (which is nonzero for only $\ell =1$ and $\ell =0$), $v \lesssim 1$ (which has contributions from all multipoles), and $v \ll 1$ (which is strongly peaked at $\ell =2$ and $\ell =0$).

While the change in the shape of the power spectrum for a given polarization mode is pronounced, it may be challenging to disentangle which modes contribute to an overall signal.
In particular, for small enough values of $v$, the significant drop in power for $\ell \neq 2$ across all spectra renders them effectively degenerate, with the exception of $C^{zz}_\ell$ for the scalar modes, which have large monopole contributions.
Moreover, the shapes of the spectra of different modes can look similar by adjusting the value of $v$, which is further complicated by the fact that the observed power spectra for subluminal GWs should have contributions from a range of velocities, corresponding to the range of observed frequencies, as discussed earlier in this section.
Regardless, the presence of any monopole or dipole contributions in PTA and astrometric measurements (barring systematic uncertainties) would be a clear indicator of physics beyond GR.

Although we may assume the GW stochastic background consists mostly, if not all, of standard GR tensor modes, we can also consider the possibility of a dominant subluminal mode that could generate a similar response to that expected from GR.
We first note that the power spectrum for standard tensor modes is dominated by the quadrupole.
While the power spectra for modes with $v \ll 1$ are also dominant at $\ell = 2$, they have a much steeper drop off at higher $\ell$.
We can instead attempt to match the GR tensor mode power spectrum more closely by considering $v \lesssim 1$, softening the drop in power at large $\ell$, but at the expense of the monopole and/or dipole contribution for non-tensor modes being more prominent.
Therefore, measurements at both small and large angular separations $\Theta$ (to probe large and small $\ell$, respectively) can help discriminate GR tensor modes from other possibilities.

\begin{figure}[t]
  \centering
  \includegraphics[width=\linewidth]{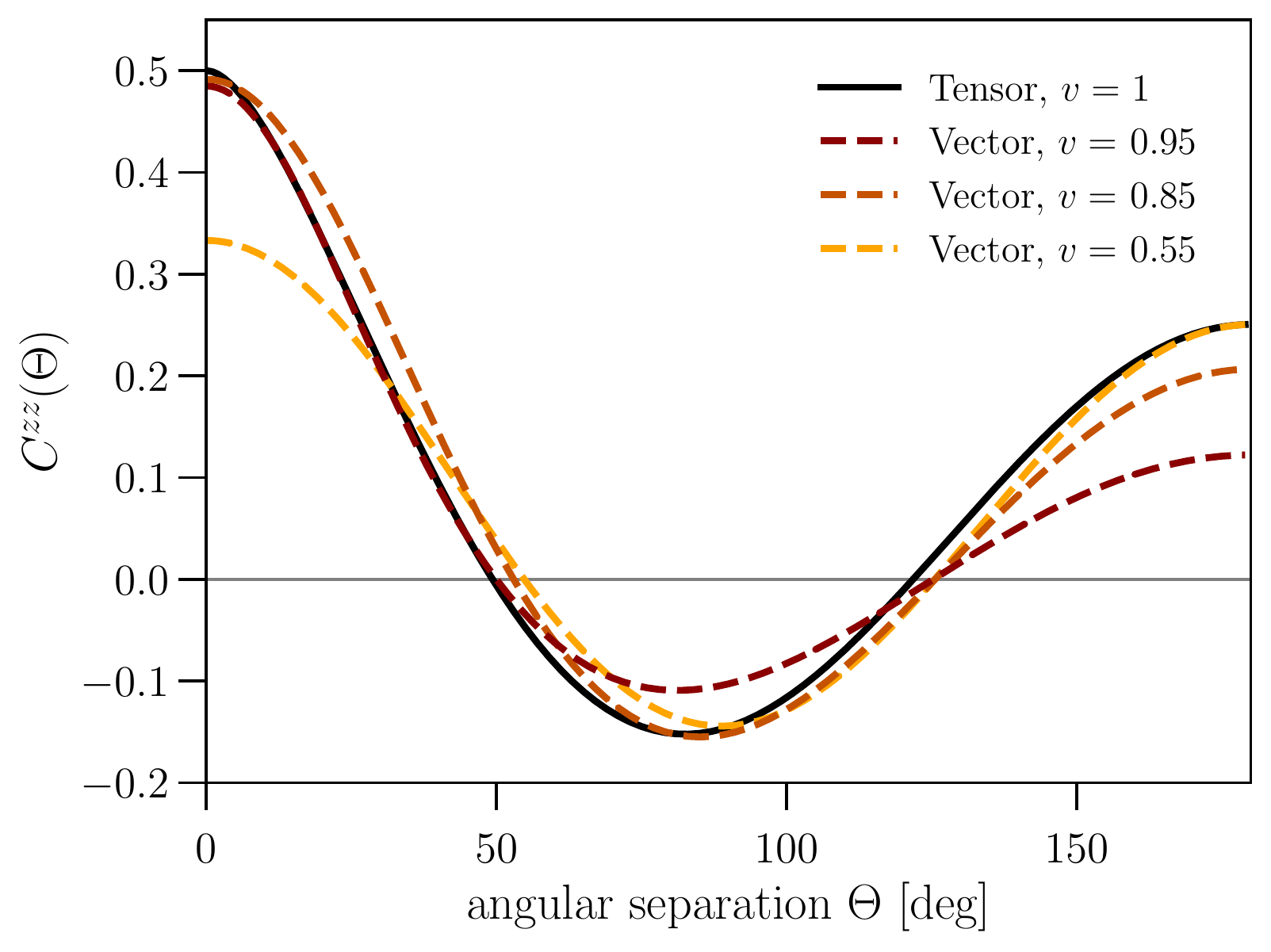}
  \caption{The $C^{zz}(\Theta)$ autocorrelation function for the luminal tensor mode (i.e. Hellings-Downs curve), as well as the vector polarization with various values of the group velocity $v$ and arbitrary normalizations chosen to resemble the Hellings-Downs curve.}
  \label{fig:corr_zz}
\end{figure}

We further demonstrate the possible similarities (or lack thereof) for PTAs by comparing the vector and tensor mode correlation functions
\begin{equation}
  C^{zz}(\Theta) = \sum_\ell \frac{2\ell +1}{4\pi}
  C_\ell^{zz} P_\ell (\cos\Theta) ,
\end{equation}
where $P_\ell$ are Legendre polynomials, at different velocities to the standard Hellings-Downs curve in Fig.~\ref{fig:corr_zz}.
Each of the vector correlation function curves are normalized separately to resemble the Hellings-Downs curve over particular ranges of $\Theta$, since any rescaling of the amplitudes cannot match the Hellings-Downs curve at both small and large angles, although the $v = 0.85$ curve comes somewhat close.
Moreover, the vector correlation functions appear slightly shifted towards larger $\Theta$ compared to the tensor modes, making measurements at many different angular separations (many multipoles) important for discrimination power.
The NANOGrav 11-year dataset, for instance, has the most sensitivity between $30^\circ$ and $60^\circ$ but has very few pulsar pairs at wide separations, above $90^\circ$~\cite{Arzoumanian:2018saf}.
Without more wide-angle pairs, low multipoles (important for non-tensor modes) cannot be well studied or constrained.

Finally, we caution that for very small $v$, our assumption that the pulsars are far compared to the wavelength of the GWs, i.e.\ the distant-source limit, ceases to be a reasonable approximation.
For GWs of a given frequency $f$, the corresponding GW wavelength is $\lambda = v_\textrm{ph} / f = 1/(fv)$; thus, decreasing $v$ increases $\lambda$.
If the wavelength of the GW is comparable to or larger than the separation between the Earth and the source or between source pairs, the distant-source limit is no longer valid~\cite{Mingarelli:2014xfa}.
For GWs of frequency $f \sim \textrm{yr}^{-1}$, there are $\sim 3\times 10^3 v$ GW wavelengths between Earth and a source located a distance $r_s \sim \textrm{kpc}$ away.
If two sources are located the same distance away, there are $\sim 10^4 v/\ell$ GW wavelengths between them for multipole $\ell$.
Thus, for the case of $v=0.01$, the distant-source limit is expected to break down for $\ell \gtrsim 10$.

%%%%%%%%%%%%%%%%%%%%%%%%%%%%%%%%%%%%%%%%%%%%%%%%%%%%%%%%%%%%%%%%%%%%%%%%%%%%%%%
\section{Gravitational Wave Polarizations in $f(R)$ gravity}
\label{sec:fR}

As an example of subluminal GW propagation, we consider $f(R)$ gravity, where $f(R)$ is a function of the Ricci scalar $R$.
The corresponding action is
\begin{equation}
  S = \frac{1}{16\pi G} \int d^4x\, \sqrt{-g} f(R) ,
  \label{eqn:action}
\end{equation}
where $G$ is the gravitational constant and $g$ is the trace of the metric $g^{\mu\nu}$.
The field equations (from varying the metric) and their trace are given by
\begin{align}
  0 &= f'(R) R_{\mu\nu} - \frac{1}{2} f(R) g_{\mu\nu}
  + \left( g_{\mu\nu} \Box - \nabla_\mu \nabla_\nu \right) f'(R) , \nonumber \\
  0 &= f'(R) R - 2 f(R) + 3 \Box f'(R) ,
  \label{eqn:field}
\end{align}
respectively, where $\Box \equiv g^{\mu\nu} \nabla_\mu \nabla_\nu$ and $R_{\mu\nu}$ is the Ricci tensor.
For the case of standard GR, $f(R)=R$.

Let us assume $f(R)$ is well-behaved in order to Taylor expand around the static vacuum value $R=0$,\footnote{For the case of $1/R^n$ gravity, where $n>0$, we would expand around a finite value $R_0$~\cite{Chiba:2006jp}.} treating $R$ as a perturbation.
To work at linear order in the expansion, we require that~\cite{Chiba:2006jp}
\begin{align}
  f(0) + f'(0) R &\gg \frac{1}{n!} f^{(n)} (0) R^n , \\
  f'(0) + f''(0) R &\gg \frac{1}{n!} f^{(n+1)} (0) R^n ,
\end{align}
for all higher-order terms with $n>1$.
We focus on $f(R)$ models that contain $R$-dependent contributions beyond GR, such that $f(0)=0$ and $f'(0) \neq 0$.
Thus, expanding $f(R)$ to linear order in $R$, Eq.~\eqref{eqn:field} becomes
\begin{align}
  0 &= m^2 \left( R_{\mu\nu} - \frac{1}{2} g_{\mu\nu} R \right) + \frac{1}{3}
  \left( g_{\mu\nu}\Box - \nabla_\mu \nabla_\nu \right) R , \nonumber\\
  0 &= \left( \Box - m^2 \right) R ,
  \label{eqn:field1}
\end{align}
with
\begin{equation}
  m^2 \equiv \frac{f'(0)}{3f''(0)} ,
  \label{eqn:mass}
\end{equation}
where we have divided out a factor of $3f''(0)$, which we assume to be nonzero.

We note that our findings are consistent with the more general statement that $f(R)$ gravity is equivalent to a scalar-tensor theory of gravity~\cite{OHanlon:1972xqa,Teyssandier:1983zz}, in which a massive scalar field $\varphi$ constitutes a single additional degree of freedom beyond GR~\cite{Liang:2017ahj}.
This connection is clear from writing the action as $S \sim \int d^4x\, \sqrt{-g} [f(\varphi) + (R-\varphi) f'(\varphi)]$.
The field equation for $\varphi$ is $\varphi = R$ if $f''(\varphi) \neq 0$, and we recover the action of Eq.~\eqref{eqn:action}.

Let us now investigate how this scalar degree of freedom decomposes into the ST and SL GW polarizations.
One can show that in synchronous gauge, the scalar perturbation can be written as
\begin{equation}
  h^{\mathrm{scalar}}_{ab} = 2 \alpha R \left( \delta_{ab} - \frac{k_a k_b}{\omega^2} \right) ,
\end{equation}
where $\alpha = 1/6m^2$, which can be interpreted as the parameter that appears in $f(R) = R + \alpha R^2$ gravity (see Appendix~\ref{sec:synchronous} for a derivation of the degrees of freedom in synchronous gauge).
For a GW propagating in the $z$ direction with $k_a = (0,0,k)$, we find
\begin{equation}
  h^{\mathrm{scalar}}_{ab} \propto
  \begin{pmatrix}
    1 & & \\
    & 1 & \\
    & & 1 - v^2
  \end{pmatrix}.
\end{equation}
The ST and SL polarization tensors are
\begin{equation}
  \varepsilon^{ST}_{ab} =
  \begin{pmatrix}
    1 & & \\
    & 1 & \\
    & & 0
  \end{pmatrix}
  \qquad
  \varepsilon^{SL}_{ab} = \sqrt{2}
  \begin{pmatrix}
    0 & & \\
    & 0 & \\
    & & 1
  \end{pmatrix},
\end{equation}
normalized to satisfy $\varepsilon^{s, ab} \varepsilon^{s'}_{ab} = 2 \delta_{ss'}$.
Thus, we find that the ratio of the SL to ST polarization amplitudes for the $f(R)$ scalar wave is $(1-v^2)/\sqrt{2} = m^2 / \sqrt{2} \omega^2$.

We can calculate the geodesic deviation, which is a gauge-invariant quantity in linearized gravity.
If we write the metric for a scalar GW as
\begin{equation}
  h^{\mathrm{scalar}}_{ab} = \left( \varepsilon^{ST}_{ab} + \frac{1 - v^2}{\sqrt{2}} \varepsilon^{SL}_{ab} \right) e^{i (kz - \omega t)} ,
  \label{eqn:fR_perturbation}
\end{equation}
then the geodesic deviation equation gives
\begin{align}
  \ddot{x} &= - \frac{\omega^2}{2} e^{i (kz - \omega t)} x \ , \\
  \ddot{y} &= - \frac{\omega^2}{2} e^{i (kz - \omega t)} y \ , \\
  \ddot{z} &= - \frac{m^2}{2} e^{i (kz - \omega t)} z.
\end{align}
This result is consistent with that in Ref.~\cite{Liang:2017ahj}, where the geodesic deviation was calculated using a different choice of gauge, and in Ref.~\cite{Moretti:2019yhs}, which used a gauge-invariant method.

The analysis in this section has thus far been in terms of plane waves, rather than TAM waves.
We can translate between the two bases using Eq.~\eqref{eqn:plane_to_TAM_tensor}.
In the case of the $f(R)$ scalar mode, the plane waves only project onto the ST and SL spherical harmonics.
Then, Eq.~\eqref{eqn:plane_to_TAM_tensor} allows us to rewrite Eq.~\eqref{eqn:fR_perturbation} as
\begin{align}
  h^{\mathrm{scalar}}_{ab} &= 4 \pi \sum_{\ell, m} i^\ell Y_{(\ell m)} (\hat{k}) \nonumber \\
  & \times \left[ \sqrt{2} \Psi^{k, ST}_{(\ell m) ab} (\boldsymbol{x}) + \left( 1 - v^2 \right) \Psi^{k, SL}_{(\ell m) ab} (\boldsymbol{x}) \right].
\end{align}
From Table~\ref{tab:F} in Appendix~\ref{sec:derive}, we note that
\begin{align}
  F_\ell^{z,ST} &= - \frac{1 - v^2}{\sqrt{2}} F_\ell^{z,SL}, \mbox{ for } \ell \geq 2 , \\
  F_\ell^{E,ST} &= - \frac{1 - v^2}{\sqrt{2}} F_\ell^{E,SL}, \mbox{ for } \ell \geq 2 ,
\end{align}
and thus the monopole and dipole are the only non-vanishing moments for the $C_\ell^{zz}$ and $C_\ell^{EE}$ power spectra for the scalar mode.
The $C_\ell^{zE}$ spectrum, however, does not experience the same cancellations and receives contributions from all multipoles.

\begin{figure}[t]
  \centering
  \includegraphics[width=\linewidth]{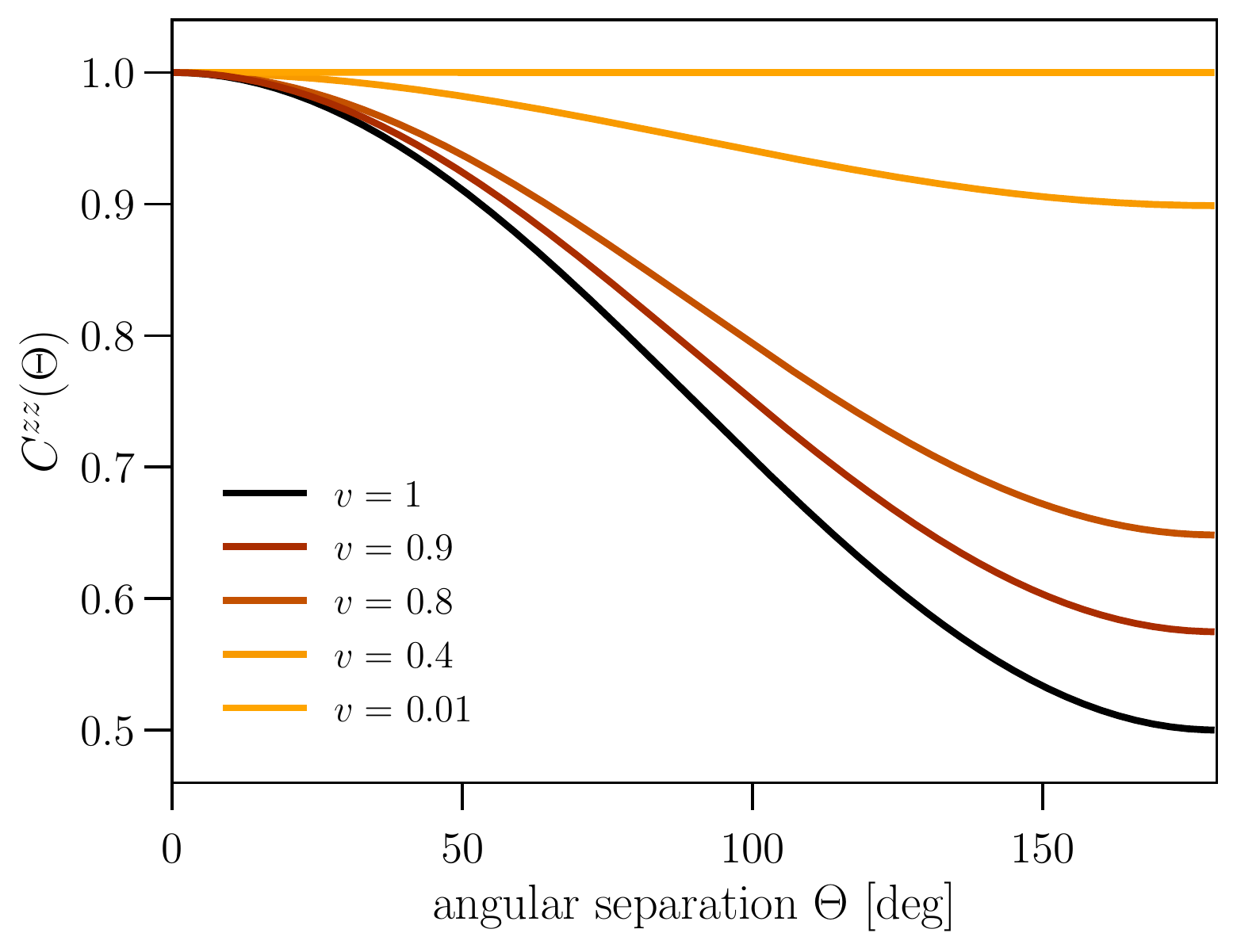}
  \caption{The $C^{zz}(\Theta)$ autocorrelation function for the $f(R)$ scalar mode at various values of the group velocity $v$.
    The correlation function only receives contributions from the monopole and dipole, the latter of which vanishes as $v \rightarrow 0$.
    Each curve is normalized such that $C^{zz}(0^\circ) = 1$.}
  \label{fig:corr_fR}
\end{figure}

Figure~\ref{fig:corr_fR} shows the redshift auto-correlation for the $f(R)$ scalar at different velocities.
As $v \rightarrow 0$, the dipole contribution vanishes, leaving only a constant correlation.
In addition, note that since the $ST$ and $SL$ polarizations are not the same as the scalar polarizations from the diagonal basis for the kinetic matrix of the theory, they have a non-vanishing cross-correlation that contributes to the $f(R)$ scalar correlations~\cite{Isi:2018miq}.
Lastly, should gravity be described by this $f(R)$ theory, we would expect the redshift angular correlation for a stochastic background of gravitational waves to look like the Hellings-Downs curve with small corrections due to this scalar correlation function.
Since any cross-correlations between the tensor and scalar modes vanish~\cite{Dai:2012bc}, the power spectrum and correlation function of the GW background should simply be a linear combination of the tensor and scalar mode contributions.

%%%%%%%%%%%%%%%%%%%%%%%%%%%%%%%%%%%%%%%%%%%%%%%%%%%%%%%%%%%%%%%%%%%%%%%%%%%%%%%
\section{Conclusions}
\label{sec:conclusions}

We have derived the power spectra for the induced time delay in pulsar-timing surveys and the induced stellar shifts in astrometry from a stochastic background of subluminal GWs.
In the limit that the GW velocity approaches the speed of light, we recover the results presented in Ref.~\cite{Qin:2018yhy}.
We have treated each GW polarization independently; however, a particular theory of modified gravity may relate the amplitudes between certain modes.
As an example, we have considered $f(R)$ gravity, which gives rise to a single massive scalar mode that is a linear combination of ST and SL modes.
The relative contribution of ST and SL is set by the group velocity $v$ of the GW.
We find that this new mode only excites the monopole and dipole.

Previous studies of the angular correlations for non-Einsteinian polarizations focused on the case of luminal GW propagation.
As there are, however, gravitational theories that contain a massive degree of freedom, our work provides the foundation for considering subluminal GW propagation in the context of stochastic GW observations.
In particular, the results of this study can be used to set bounds on subluminal GWs from the low-frequency end of the spectrum.
We also find that our results numerically agree with those in Ref.~\cite{Mihaylov:2019lft} for subluminal GWs.
Moreover, while we are able to obtain analytic expressions for the power spectra, we do not have analytic expressions for the correlation functions; Ref.~\cite{Mihaylov:2019lft} has analytic expressions for the correlation functions, but the power spectra must be numerically computed.
Our results and these other results are thus complementary.
If PTAs or astrometric surveys find evidence of correlations beyond what is expected from GR, we may use our findings to account for the effects subluminal propagation that may arise in particular models.

%%%%%%%%%%%%%%%%%%%%%%%%%%%%%%%%%%%%%%%%%%%%%%%%%%%%%%%%%%%%%%%%%%%%%%%%%%%%%%%

\begin{acknowledgments}
We thank Chiara Mingarelli and Tristan Smith for their comments on the manuscript.
We also thank Sylvia Biscoveanu, F\'{e}lix-Louis Juli\'{e}, and Ryan McManus for helpful conversations.
WQ acknowledges the support of a Space@Hopkins grant, the MIT Department of Physics, and the NSF GRFP.
MK acknowledges the support of NASA Grant No.\ NNX17AK38G, NSF Grant No.\ 0244990, and the Simons Foundation.
\end{acknowledgments}

%%%%%%%%%%%%%%%%%%%%%%%%%%%%%%%%%%%%%%%%%%%%%%%%%%%%%%%%%%%%%%%%%%%%%%%%%%%%%%%

\appendix

\section{Projection factors}
\label{sec:derive}

In this Appendix, we outline the analytic calculation of the projection factors $F_\ell^{z,\alpha}$, $F_\ell^{E,\alpha}$, and $F_\ell^{B,\alpha}$ defined in Eqs.~\eqref{eqn:Fz}, \eqref{eqn:FE}, and \eqref{eqn:FB}, respectively, for each GW polarization $\alpha$.
These equations rely on the radial functions given explicitly in Refs.~\cite{Dai:2012bc,Qin:2018yhy}.
However, directly applying these functions as written using Mathematica yields cumbersome expressions for the projection factors that are difficult to relate to our previous work.

In order to obtain clean expressions, we first need to rewrite the radial functions in a form that is amenable to calculating the projection factors by hand:
\begin{widetext}
\begin{align*}
  R_\ell^{L,SL}(x) &= j^{\prime\prime}_\ell(x) &
  R_\ell^{E,SL}(x) &= -\frac{x}{2} \frac{d}{dx}R_\ell^{E,SL}(x) + \frac{1}{2}\sqrt{\ell(\ell+1)} R_\ell^{L,SL}(x) \\
  R_\ell^{L,ST}(x) &= -\frac{1}{\sqrt{2}} \left[R_\ell^{L,SL}(x) + j_\ell(x)\right] &
  R_\ell^{E,ST}(x) &= -\frac{1}{\sqrt{2}} R_\ell^{E,SL}(x) \\
  R_\ell^{L,VE}(x) &= -\sqrt{2\ell(\ell+1)} \frac{d}{dx} \left[\frac{j_\ell(x)}{x} \right] &
  R_\ell^{E,VE}(x) &= -\frac{x}{2} \frac{d}{dx}R_\ell^{E,VE}(x) + \frac{1}{2}\sqrt{\ell(\ell+1)} R_\ell^{L,VE}(x) + \frac{x}{2\sqrt{2}} j'_\ell(x) \\
  R_\ell^{L,TE}(x) &= -N_\ell \frac{j_\ell(x)}{x^2} &
  R_\ell^{E,TE}(x) &= -\frac{x}{2} \frac{d}{dx}R_\ell^{E,TE}(x) + \frac{1}{2}\sqrt{\ell(\ell+1)} R_\ell^{L,TE}(x) + \frac{N_\ell}{2\sqrt{\ell(\ell+1)}} j_\ell(x) \\
  && R_\ell^{B,VB}(x) &= \frac{ix}{2\sqrt{\ell(\ell+1)}} R_\ell^{L,VE}(x) \\
  && R_\ell^{B,TB}(x) &= \frac{ix}{\sqrt{\ell(\ell+1)}} R_\ell^{L,TE}(x) ,
\end{align*}
\end{widetext}
where $j_\ell(x)$ is the spherical Bessel function of the first kind, $N_\ell \equiv \sqrt{(\ell+2)!/[2(\ell-2)!]}$ , and functions with unlisted combinations of $\{L,E,B\}$ and $\alpha$ are zero.
Note that we have used the differential equation for the spherical Bessel function
\begin{equation}
  x^2 j^{\prime\prime}_\ell(x) + 2x j'_\ell(x) + \left[x^2-\ell(\ell+1)\right] j_\ell(x) = 0
\end{equation}
to recast $R_\ell^{L,SL}$ from its form given in Refs.~\cite{Dai:2012bc,Qin:2018yhy}.
The simple relations between the radial functions of the ST and SL modes allow for the projection factors for ST to be easily obtained from those for SL.
Additionally, $F_\ell^{B,VB}$ and $F_\ell^{B,TB}$ are easily obtained from $F_\ell^{z,VE}$ and $F_\ell^{z,TE}$, respectively.
The somewhat complicated relations for $R_\ell^{E,\alpha}$ are particularly useful to simplify the integrand in Eq.~\eqref{eqn:FE}.

Plugging these radial functions into the equations for the projection factors, we simplify expression by integrating by parts any function with an explicit derivative.
All boundary terms are proportional to $j_\ell(x)/x^n$ or $j'_\ell(x)/x^n$ for $n\geq 0$, which vanish at $x\to\infty$; terms evaluated at $x\to 0$ are determined by the limiting behavior $j_\ell(x) \to x^\ell 2^{-(\ell+1)} \sqrt{\pi}/\Gamma(\ell+3/2)$.
The remaining terms in the projection factors are all proportional to
\begin{equation}
  I_\ell^{(n)}(v) \equiv \int_0^\infty \frac{j_\ell(x)}{x^n} e^{ix\vp} \, dx
  \label{eqn:Igen}
\end{equation}
with $v = 1/\vp$.
For instance, we determine $F_\ell^{z,SL}$ by integrating $j^{\prime\prime}_\ell(x)$ by parts twice, leaving a term proportional to $I_\ell^{(0)}$; meanwhile, the boundary terms at $x\to 0$ involve $j_\ell \to \delta_{\ell 0}$ and $j'_\ell \to \delta_{\ell 1}/3$.
We summarize our results in Table~\ref{tab:F}.

\begin{table*}
  \centering
  \everymath{\displaystyle}
  \resizebox{0.95\textwidth}{!}{%
    \begin{tabular}{|Sc|Sc|Sc|Sc|}
      \hline
      $\alpha$ & $F^{E,\alpha}_\ell$ & $F^{B,\alpha}_\ell$ & $F^{z,\alpha}_\ell$ \\
      \hline \hline
      $ST$ & $\frac{i}{6 v} \delta_{\ell 1} - \frac{1-v^2}{2 v^2}\sqrt{\frac{\ell(\ell+1)}{2}} I_\ell^{(1)}(v)$ & 0
      & $-\frac{1}{2\sqrt{2}} \left[\frac{1}{v^2} \delta_{\ell 0} + \frac{i}{3 v} \delta_{\ell 1} + i \frac{1-v^2}{v^3} I_\ell^{(0)}(v) \right]$ \\
      \hline
      $SL$ & $-\frac{i}{3\sqrt{2} v} \delta_{\ell 1} + \frac{1}{2 v^2}\sqrt{\ell(\ell+1)} I_\ell^{(1)}(v)$ & 0
      & $\frac{1}{2}\left(\frac{1}{v^2} \delta_{\ell 0} + \frac{i}{3v}\delta_{\ell 1}\right) + \frac{i}{2 v^3} I_\ell^{(0)}(v)$ \\
      \hline
      $VE$ &$\frac{2i}{3\sqrt{2} v} \delta_{\ell 1} - \frac{1}{\sqrt{2} v^2} I_\ell^{(1)}(v) + \frac{i (1-v^2)}{\sqrt{2} v^3} I_\ell^{(0)}(v)$ & 0
      & $-\frac{i}{3 v} \delta_{\ell 1} + \frac{1}{v^2} \sqrt{\frac{\ell(\ell+1)}{2}} I_\ell^{(1)}(v)$ \\
      \hline
      $VB$ & 0 & $\frac{i}{3\sqrt{2}} \delta_{\ell 1} - \frac{1}{\sqrt{2} v} I_\ell^{(1)}(v)$ & 0 \\
      \hline
      $TE$ & $-\frac{N_\ell}{\sqrt{\ell(\ell+1)}} \left[\frac{i}{v} I_\ell^{(2)}(v) + \frac{1-v^2}{2 v^2} I_\ell^{(1)}(v) \right]$ & 0
      & $\frac{i}{2v} N_\ell I_\ell^{(2)}(v)$ \\
      \hline
      $TB$ & 0 & $-\frac{iN_\ell}{\sqrt{\ell (\ell+1)}} I_\ell^{(2)}(v)$ & 0 \\
      \hline
    \end{tabular}
  }
  \caption{Projection factors defined in Eqs.~\eqref{eqn:Fz}, \eqref{eqn:FE}, and \eqref{eqn:FB} that relate the amplitude of a given TAM wave to its associated observables.
    The first column $\alpha$ labels the GW polarization.
    Note that $v$ is the group velocity, related to the phase velocity by $v=1/\vp$ for massive gravity.
    We define $N_\ell \equiv \sqrt{(\ell+2)!/[2(\ell-2)!]}$ and define $I_\ell^{(n)}$ in Eq.~\eqref{eqn:Igen}.}
  \label{tab:F}
\end{table*}

Up to this point, all derivations and the results in Table~\ref{tab:F} hold for generic $v$.
If we fix $v=1$, Eq.~\eqref{eqn:Igen} becomes
\begin{equation}
  I_\ell^{(n)}(v=1) = i^{\ell+1-n}\, 2^{n-1} \frac{(\ell-n)!}{(\ell+n)!} (n-1)!
  \label{eqn:Ilum}
\end{equation}
for $\ell+1 > n > 0$.
Table~\ref{tab:F} then matches the analogous table in our previous work~\cite{Qin:2018yhy}.
Note that Eq.~\eqref{eqn:Ilum} diverges for $n=0$, corresponding to the divergence of $F_\ell^{z,SL}$ discussed in Ref.~\cite{Qin:2018yhy}.
Other terms in Table~\ref{tab:F} involving $n=0$ should be dropped due to $(1-v^2)$ prefactors in order to recover previous results.

For the case of subluminal GWs, we have
\begin{align}
  I_\ell^{(n)}(v<1) =& \left(\frac{i v}{2}\right)^{\ell+1-n} \frac{\sqrt{\pi}}{2^n} \frac{\Gamma(\ell+1-n)}{\Gamma(\ell + \frac{3}{2})} \label{eqn:Ifhg} \\
  & \times \fhg \left(\frac{\ell+1-n}{2},\frac{\ell+2-n}{2},\ell+\frac{3}{2},v^{2} \right) \nonumber
\end{align}
for $\ell + 1 > n \geq 0$, where $\fhg$ is Gauss's hypergeometric function.
We have verified numerically that our results for the power spectra and correlation functions agree with those in Ref.~\cite{Mihaylov:2019lft}.

As $v \rightarrow 0$, the hypergeometric function in Eq.~\eqref{eqn:Ifhg} approaches 1 at leading order and $I_\ell^{(n)}(v) \sim v^{\ell+1-n}$.
From Table~\ref{tab:F}, all terms for $F_\ell^{E,\alpha}$ and $F_\ell^{z,\alpha}$ that involve $I_\ell^{(n)}(v)$ thus have a $v^{\ell -2}$ dependence in this limit; for $F_\ell^{B,VB}$ and $F_\ell^{B,TB}$, the velocity dependence of the $I_\ell^{(n)}(v)$ terms is $v^{\ell -1}$.
Therefore, the projection factors for all polarizations with $\ell > 2$ are more suppressed compared to $\ell =2$ for smaller $v$, and $F_2^{E,\alpha}$ and $F_2^{z,\alpha}$ approach constant values.
As expected from the discussion in Sec.~\ref{sec:power_spectra}, all of our power spectra feature a dominate quadrupole, seen in Figs.~\ref{fig:czz}-\ref{fig:cze}.

In order to consider the $v \rightarrow 0$ limit for projection factors with $\ell=0$ or $1$, we must account for cancellations between the $I_\ell^{(n)}(v)$ terms and any $\delta_{\ell 0}$ or $\delta_{\ell 1}$ terms and use the expansion $\fhg(a,b,c,v^2) \rightarrow 1+(ab/c)v^2$.
For all relevant (i.e., scalar and vector) projection factors, the leading order contribution for $\ell=1$ scales with one additional power of $v$ compared to the $\ell=2$ case, resulting in the suppression of the dipole with respect to the quadrupole in Figs.~\ref{fig:czz}-\ref{fig:cze}.
Finally, $F_\ell^{z,ST}$ and $F_\ell^{z,SL}$ approach constant values for $\ell =0$, with $F_0^{z,ST} / F_2^{z,ST} \rightarrow 5$ and $F_0^{z,SL} / F_2^{z,SL} \rightarrow -5/2$.
Therefore, the $C_\ell^{zz}$ power spectra for the ST and SL modes exhibit monopole contributions that are factors of $25$ and $25/4$ larger than the quadrupole, respectively, as observed in Fig.~\ref{fig:czz}.

Although we do not discuss the possibility of superluminal propagation in this work, we can apply Table~\ref{tab:F} to such a scenario, writing Eq.~\eqref{eqn:Igen} as
\begin{widetext}
\begin{align}
  I_\ell^{(n)}(v>1) = 2^{-(n+1)} \sqrt{\pi} & \left\{
    \frac{\Gamma\left(\frac{\ell+1-n}{2}\right)}
         {\Gamma\left(\frac{\ell+2+n}{2}\right)}
    \fhg\left(-\frac{\ell+n}{2},\frac{\ell+1-n}{2},\frac{1}{2},v^{-2}\right)
    - \frac{2iv}{1+2\ell}
    \frac{\Gamma\left(\frac{\ell+2-n}{2}\right)}
         {\Gamma\left(\frac{\ell+1+n}{2}\right)}
    \times \right. \nonumber\\
    &\left. \times \left[
      \fhg\left(-\frac{\ell+1+n}{2},\frac{\ell+2-n}{2},\frac{1}{2},v^{-2}\right)
      - \fhg\left(-\frac{\ell-1+n}{2},\frac{\ell-n}{2},\frac{1}{2},v^{-2}\right)
      \right]
    \right\}
\end{align}
\end{widetext}
for $\ell + 1 > n \geq 0$.

%%%%%%%%%%%%%%%%%%%%%%%%%%%%%%%%%%%%%%%%
\section{Synchronous gauge}
\label{sec:synchronous}

To determine the degrees of freedom in $f(R)$ gravity from the field equations, we start with a generic metric perturbation with components
\begin{align}
  h_{00} &= -2 \Phi , \nonumber \\
  h_{0a} &= w_a , \nonumber \\
  h_{ab} &= 2 s_{ab} - 2 \Psi \delta_{ab} ,
\end{align}
where $\Psi = - \frac{1}{6} \delta^{ab} h_{ab}$ is proportional to the trace and $s_{ab} = \frac{1}{2} \left( h_{ab} - \frac{1}{3} \delta^{cd} h_{cd} \delta_{ab} \right)$ is traceless.
From Eq.~\eqref{eqn:field1}, we already know that this theory has a propagating scalar degree of freedom, $R$, so we will eventually rewrite some of the above components in terms of $R$.

\begin{widetext}
  The Ricci tensor for this metric is
  \begin{align}
    R_{00} &= \nabla^2 \Phi + \partial_0 \partial_c w^c + 3 \partial_0^2 \Psi , \nonumber \\
    R_{0b} &= - \frac{1}{2} \nabla^2 w_b + \frac{1}{2} \partial_b \partial_c w^c + 2 \partial_0 \partial_b \Psi + \partial_0 \partial_c s_b^c , \nonumber \\
    R_{ab} &= - \partial_a \partial_b (\Phi - \Psi) - \partial_0 \partial_{(a} w_{b)} + \Box \Psi \delta_{ab} - \Box s_{ab} + 2 \partial_c \partial_{(a} s_{b)}^c ,
  \end{align}
  and the Ricci scalar is therefore
  \begin{equation}
    R = - R_{00} + R_{aa} = -2 \nabla^2 \Phi - 6 \partial_0^2 \Psi + 4 \nabla^2 \Psi -2 \partial_0 \partial_a w^a + 2 \partial_c \partial_{a} s^{ac} .
    \label{eqn:general_scalar}
  \end{equation}
  Substituting these results into the field equations, we find for the various components:
  \begin{align}
    00: \quad 0&= m^2 \left( \nabla^2 \Phi + \partial_0 \partial_c w^c + 3 \partial_0^2 \Psi + \frac{1}{3} R \right) + \frac{1}{6} (m^2 - \nabla^2) R , \\
    0a: \quad 0&= m^2 \left( - \frac{1}{2} \nabla^2 w_a + \frac{1}{2} \partial_a \partial_b w^b + 2 \partial_0 \partial_a \Psi + \partial_0 \partial_b s_a^b \right) - \frac{1}{6} \partial_0 \partial_a R , \\
    ab: \quad 0&= m^2 \left[- \partial_a \partial_b (\Phi - \Psi) - \partial_0 \partial_{(a} w_{b)} + \Box \Psi \delta_{ab} - \Box s_{ab} + 2 \partial_c \partial_{(a} s_{b)}^c - \frac{1}{3} \delta_{ab} R \right] - \frac{1}{6} \partial_a \partial_b R .
  \end{align}

  The first equation has no time derivatives of $\Phi$ and the second has no time derivatives of $w_a$, so these two fields do not represent propagating degrees of freedom---they can be written purely in terms of other fields.
  We work in the synchronous gauge by setting $\Phi = w^a = 0$.
  The field equations simplify to
  \begin{align}
    00: \quad 0&= m^2 \left( 3 \partial_0^2 \Psi + \frac{1}{3} R \right) + \frac{1}{6} (m^2 - \nabla^2) R , \\
    0a: \quad 0&= m^2 \left( 2 \partial_0 \partial_a \Psi + \partial_0 \partial_b s_a^b \right) - \frac{1}{6} \partial_0 \partial_a R , \\
    ab: \quad 0&= m^2 \left[ \partial_a \partial_b \Psi + \Box \Psi \delta_{ab} - \Box s_{ab} + 2 \partial_c \partial_{(a} s_{b)}^c - \frac{1}{3} \delta_{ab} R \right] - \frac{1}{6} \partial_a \partial_b R .
  \end{align}
\end{widetext}

The $00$ equation allows us to write $\Psi$ purely in terms of $R$.
Since $R$ satisfies a wave equation, let us assume $\partial_a R = ik_a R$ and $\partial_0 R = -i \omega R$, where $\omega^2 - k^2 = m^2$.
If we set the time-independent integration constants to zero, the $00$ field equation can be integrated to give
\begin{equation}
  \Psi = \frac{3 \omega^2 - 2 k^2}{18 m^2 \omega^2} R.
\end{equation}
Similarly, integrating the $0a$ equation with respect to time, we find
\begin{align}
  \partial^b s_{ab} = \partial_a \left( \frac{1}{6 m^2} R - 2 \Psi \right) \propto R.
\end{align}
Thus, $\Psi$ and the divergence of $s_{ab}$ are both related to the same single degree of freedom: the scalar $R$.

The only equation we have not yet studied is the $ab$ equation.
Substituting $\Psi$ and $\partial^b s_{ab}$ with their relations to $R$, we find the $ab$ equation simplifies to
\begin{equation}
  \Box s_{ab} = \frac{6 k_a k_b - \left( 2 k^2 + 3 \omega^2 \right) \delta_{ab}}{18 \omega^2} R .
\end{equation}
Then, since $R = \frac{\Box}{m^2} R$, we can rewrite this as
\begin{equation}
  \Box \left[ s_{ab} + \frac{\left( 2 k^2 + 3 \omega^2 \right) \delta_{ab} - 6 k_a k_b}{3 \omega^2} \,\alpha R \right] = 0 ,
  \label{eqn:tensor_dofs}
\end{equation}
where $\alpha = 1/6m^2$.

The full metric perturbation is $h_{ab} = 2 s_{ab} - 2 \Psi \delta_{ab}$.
If the tensor degrees of freedom are of the form $s_{ab} + C_{ab} R$, where $C_{ab}$ denotes the coefficient of $R$ in Eq.~\eqref{eqn:tensor_dofs}, the remaining scalar perturbation becomes
\begin{equation}
  h^{\mathrm{scalar}}_{ab} \equiv \Psi \delta_{ab} + C_{ab} R = 2 \alpha R \left( \delta_{ab} - \frac{k_a k_b}{\omega^2} \right) .
\end{equation}

Thus, after fixing the synchronous gauge, we have shown that our remaining degrees of freedom are a massive scalar $R$ and two tensor degrees of freedom that satisfy a massless wave equation.

%%%%%%%%%%%%%%%%%%%%%%%%%%%%%%%%%%%%%%%%%%%%%%%%%%%%%%%%%%%%%%%%%%%%%%%%%%%%%%%

\bibliography{subluminal}

\begin{thebibliography}{36}%
\makeatletter
\providecommand \@ifxundefined [1]{%
 \@ifx{#1\undefined}
}%
\providecommand \@ifnum [1]{%
 \ifnum #1\expandafter \@firstoftwo
 \else \expandafter \@secondoftwo
 \fi
}%
\providecommand \@ifx [1]{%
 \ifx #1\expandafter \@firstoftwo
 \else \expandafter \@secondoftwo
 \fi
}%
\providecommand \natexlab [1]{#1}%
\providecommand \enquote  [1]{``#1''}%
\providecommand \bibnamefont  [1]{#1}%
\providecommand \bibfnamefont [1]{#1}%
\providecommand \citenamefont [1]{#1}%
\providecommand \href@noop [0]{\@secondoftwo}%
\providecommand \href [0]{\begingroup \@sanitize@url \@href}%
\providecommand \@href[1]{\@@startlink{#1}\@@href}%
\providecommand \@@href[1]{\endgroup#1\@@endlink}%
\providecommand \@sanitize@url [0]{\catcode `\\12\catcode `\$12\catcode
  `\&12\catcode `\#12\catcode `\^12\catcode `\_12\catcode `\%12\relax}%
\providecommand \@@startlink[1]{}%
\providecommand \@@endlink[0]{}%
\providecommand \url  [0]{\begingroup\@sanitize@url \@url }%
\providecommand \@url [1]{\endgroup\@href {#1}{\urlprefix }}%
\providecommand \urlprefix  [0]{URL }%
\providecommand \Eprint [0]{\href }%
\providecommand \doibase [0]{http://dx.doi.org/}%
\providecommand \selectlanguage [0]{\@gobble}%
\providecommand \bibinfo  [0]{\@secondoftwo}%
\providecommand \bibfield  [0]{\@secondoftwo}%
\providecommand \translation [1]{[#1]}%
\providecommand \BibitemOpen [0]{}%
\providecommand \bibitemStop [0]{}%
\providecommand \bibitemNoStop [0]{.\EOS\space}%
\providecommand \EOS [0]{\spacefactor3000\relax}%
\providecommand \BibitemShut  [1]{\csname bibitem#1\endcsname}%
\let\auto@bib@innerbib\@empty
%</preamble>
\bibitem [{\citenamefont {Verbiest}\ \emph {et~al.}(2016)\citenamefont
  {Verbiest} \emph {et~al.}}]{Verbiest:2016vem}%
  \BibitemOpen
  \bibfield  {author} {\bibinfo {author} {\bibfnamefont {J.~P.~W.}\
  \bibnamefont {Verbiest}} \emph {et~al.},\ }\bibfield  {title} {\enquote
  {\bibinfo {title} {{The International Pulsar Timing Array: First Data
  Release}},}\ }\href {\doibase 10.1093/mnras/stw347} {\bibfield  {journal}
  {\bibinfo  {journal} {Mon. Not. Roy. Astron. Soc.}\ }\textbf {\bibinfo
  {volume} {458}},\ \bibinfo {pages} {1267--1288} (\bibinfo {year} {2016})},\
  \Eprint {http://arxiv.org/abs/1602.03640} {arXiv:1602.03640 [astro-ph.IM]}
  \BibitemShut {NoStop}%
%%CITATION = ARXIV:1602.03640;%%
\bibitem [{\citenamefont {Perera}\ \emph {et~al.}(2019)\citenamefont {Perera}
  \emph {et~al.}}]{Perera:2019sca}%
  \BibitemOpen
  \bibfield  {author} {\bibinfo {author} {\bibfnamefont {B.~B.~P.}\
  \bibnamefont {Perera}} \emph {et~al.},\ }\bibfield  {title} {\enquote
  {\bibinfo {title} {{The International Pulsar Timing Array: Second data
  release}},}\ }\href {\doibase 10.1093/mnras/stz2857} {\bibfield  {journal}
  {\bibinfo  {journal} {Mon. Not. Roy. Astron. Soc.}\ }\textbf {\bibinfo
  {volume} {490}},\ \bibinfo {pages} {4666--4687} (\bibinfo {year} {2019})},\
  \Eprint {http://arxiv.org/abs/1909.04534} {arXiv:1909.04534 [astro-ph.HE]}
  \BibitemShut {NoStop}%
%%CITATION = ARXIV:1909.04534;%%
\bibitem [{\citenamefont {Hobbs}(2013)}]{Hobbs:2013aka}%
  \BibitemOpen
  \bibfield  {author} {\bibinfo {author} {\bibfnamefont {G.}~\bibnamefont
  {Hobbs}},\ }\bibfield  {title} {\enquote {\bibinfo {title} {{The Parkes
  Pulsar Timing Array}},}\ }\href {\doibase 10.1088/0264-9381/30/22/224007}
  {\bibfield  {journal} {\bibinfo  {journal} {Class. Quant. Grav.}\ }\textbf
  {\bibinfo {volume} {30}},\ \bibinfo {pages} {224007} (\bibinfo {year}
  {2013})},\ \Eprint {http://arxiv.org/abs/1307.2629} {arXiv:1307.2629
  [astro-ph.IM]} \BibitemShut {NoStop}%
%%CITATION = ARXIV:1307.2629;%%
\bibitem [{\citenamefont {Manchester}\ \emph {et~al.}(2013)\citenamefont
  {Manchester} \emph {et~al.}}]{Manchester:2012za}%
  \BibitemOpen
  \bibfield  {author} {\bibinfo {author} {\bibfnamefont {R.~N.}\ \bibnamefont
  {Manchester}} \emph {et~al.},\ }\bibfield  {title} {\enquote {\bibinfo
  {title} {{The Parkes Pulsar Timing Array Project}},}\ }\href {\doibase
  10.1017/pasa.2012.017} {\bibfield  {journal} {\bibinfo  {journal} {Publ.
  Astron. Soc. Austral.}\ }\textbf {\bibinfo {volume} {30}},\ \bibinfo {pages}
  {17} (\bibinfo {year} {2013})},\ \Eprint {http://arxiv.org/abs/1210.6130}
  {arXiv:1210.6130 [astro-ph.IM]} \BibitemShut {NoStop}%
%%CITATION = ARXIV:1210.6130;%%
\bibitem [{\citenamefont {Lentati}\ \emph {et~al.}(2015)\citenamefont {Lentati}
  \emph {et~al.}}]{Lentati:2015qwp}%
  \BibitemOpen
  \bibfield  {author} {\bibinfo {author} {\bibfnamefont {L.}~\bibnamefont
  {Lentati}} \emph {et~al.},\ }\bibfield  {title} {\enquote {\bibinfo {title}
  {{European Pulsar Timing Array Limits On An Isotropic Stochastic
  Gravitational-Wave Background}},}\ }\href {\doibase 10.1093/mnras/stv1538}
  {\bibfield  {journal} {\bibinfo  {journal} {Mon. Not. Roy. Astron. Soc.}\
  }\textbf {\bibinfo {volume} {453}},\ \bibinfo {pages} {2576--2598} (\bibinfo
  {year} {2015})},\ \Eprint {http://arxiv.org/abs/1504.03692} {arXiv:1504.03692
  [astro-ph.CO]} \BibitemShut {NoStop}%
%%CITATION = ARXIV:1504.03692;%%
\bibitem [{\citenamefont {Hellings}\ and\ \citenamefont
  {Downs}(1983)}]{Hellings:1983fr}%
  \BibitemOpen
  \bibfield  {author} {\bibinfo {author} {\bibfnamefont {R.~w.}\ \bibnamefont
  {Hellings}}\ and\ \bibinfo {author} {\bibfnamefont {G.~s.}\ \bibnamefont
  {Downs}},\ }\bibfield  {title} {\enquote {\bibinfo {title} {{UPPER LIMITS ON
  THE ISOTROPIC GRAVITATIONAL RADIATION BACKGROUND FROM PULSAR TIMING
  ANALYSIS}},}\ }\href {\doibase 10.1086/183954} {\bibfield  {journal}
  {\bibinfo  {journal} {Astrophys. J.}\ }\textbf {\bibinfo {volume} {265}},\
  \bibinfo {pages} {L39--L42} (\bibinfo {year} {1983})}\BibitemShut {NoStop}%
%%CITATION = ASJOA,265,L39;%%
\bibitem [{\citenamefont {Arzoumanian}\ \emph {et~al.}(2020)\citenamefont
  {Arzoumanian} \emph {et~al.}}]{Arzoumanian:2020vkk}%
  \BibitemOpen
  \bibfield  {author} {\bibinfo {author} {\bibfnamefont {Zaven}\ \bibnamefont
  {Arzoumanian}} \emph {et~al.} (\bibinfo {collaboration} {NANOGrav}),\
  }\bibfield  {title} {\enquote {\bibinfo {title} {{The NANOGrav 12.5-year Data
  Set: Search For An Isotropic Stochastic Gravitational-Wave Background}},}\
  }\href@noop {} {\  (\bibinfo {year} {2020})},\ \Eprint
  {http://arxiv.org/abs/2009.04496} {arXiv:2009.04496 [astro-ph.HE]}
  \BibitemShut {NoStop}%
\bibitem [{\citenamefont {Braginsky}\ \emph {et~al.}(1990)\citenamefont
  {Braginsky}, \citenamefont {Kardashev}, \citenamefont {Novikov},\ and\
  \citenamefont {Polnarev}}]{Braginsky:1989pv}%
  \BibitemOpen
  \bibfield  {author} {\bibinfo {author} {\bibfnamefont {V.~B.}\ \bibnamefont
  {Braginsky}}, \bibinfo {author} {\bibfnamefont {N.~S.}\ \bibnamefont
  {Kardashev}}, \bibinfo {author} {\bibfnamefont {I.~D.}\ \bibnamefont
  {Novikov}}, \ and\ \bibinfo {author} {\bibfnamefont {A.~G.}\ \bibnamefont
  {Polnarev}},\ }\bibfield  {title} {\enquote {\bibinfo {title} {{Propagation
  of electromagnetic radiation in a random field of gravitational waves and
  space radio interferometry}},}\ }\href@noop {} {\bibfield  {journal}
  {\bibinfo  {journal} {Nuovo Cim.}\ }\textbf {\bibinfo {volume} {B105}},\
  \bibinfo {pages} {1141--1158} (\bibinfo {year} {1990})}\BibitemShut {NoStop}%
%%CITATION = NUCIA,B105,1141;%%
\bibitem [{\citenamefont {Kaiser}\ and\ \citenamefont
  {Jaffe}(1997)}]{Kaiser:1996wk}%
  \BibitemOpen
  \bibfield  {author} {\bibinfo {author} {\bibfnamefont {Nick}\ \bibnamefont
  {Kaiser}}\ and\ \bibinfo {author} {\bibfnamefont {Andrew~H.}\ \bibnamefont
  {Jaffe}},\ }\bibfield  {title} {\enquote {\bibinfo {title} {{Bending of light
  by gravity waves}},}\ }\href {\doibase 10.1086/304357} {\bibfield  {journal}
  {\bibinfo  {journal} {Astrophys. J.}\ }\textbf {\bibinfo {volume} {484}},\
  \bibinfo {pages} {545--554} (\bibinfo {year} {1997})},\ \Eprint
  {http://arxiv.org/abs/astro-ph/9609043} {arXiv:astro-ph/9609043 [astro-ph]}
  \BibitemShut {NoStop}%
%%CITATION = ASTRO-PH/9609043;%%
\bibitem [{\citenamefont {Book}\ and\ \citenamefont
  {Flanagan}(2011)}]{Book:2010pf}%
  \BibitemOpen
  \bibfield  {author} {\bibinfo {author} {\bibfnamefont {Laura~G.}\
  \bibnamefont {Book}}\ and\ \bibinfo {author} {\bibfnamefont {Eanna~E.}\
  \bibnamefont {Flanagan}},\ }\bibfield  {title} {\enquote {\bibinfo {title}
  {{Astrometric Effects of a Stochastic Gravitational Wave Background}},}\
  }\href {\doibase 10.1103/PhysRevD.83.024024} {\bibfield  {journal} {\bibinfo
  {journal} {Phys. Rev.}\ }\textbf {\bibinfo {volume} {D83}},\ \bibinfo {pages}
  {024024} (\bibinfo {year} {2011})},\ \Eprint {http://arxiv.org/abs/1009.4192}
  {arXiv:1009.4192 [astro-ph.CO]} \BibitemShut {NoStop}%
%%CITATION = ARXIV:1009.4192;%%
\bibitem [{\citenamefont {{Lee}}\ \emph {et~al.}(2008)\citenamefont {{Lee}},
  \citenamefont {{Jenet}},\ and\ \citenamefont {{Price}}}]{Lee:2008aa}%
  \BibitemOpen
  \bibfield  {author} {\bibinfo {author} {\bibfnamefont {K.~J.}\ \bibnamefont
  {{Lee}}}, \bibinfo {author} {\bibfnamefont {F.~A.}\ \bibnamefont {{Jenet}}},
  \ and\ \bibinfo {author} {\bibfnamefont {R.~H.}\ \bibnamefont {{Price}}},\
  }\bibfield  {title} {\enquote {\bibinfo {title} {{Pulsar Timing as a Probe of
  Non-Einsteinian Polarizations of Gravitational Waves}},}\ }\href {\doibase
  10.1086/591080} {\bibfield  {journal} {\bibinfo  {journal} {\apj}\ }\textbf
  {\bibinfo {volume} {685}},\ \bibinfo {pages} {1304--1319} (\bibinfo {year}
  {2008})}\BibitemShut {NoStop}%
\bibitem [{\citenamefont {Chamberlin}\ and\ \citenamefont
  {Siemens}(2012)}]{Chamberlin:2011ev}%
  \BibitemOpen
  \bibfield  {author} {\bibinfo {author} {\bibfnamefont {Sydney~J.}\
  \bibnamefont {Chamberlin}}\ and\ \bibinfo {author} {\bibfnamefont {Xavier}\
  \bibnamefont {Siemens}},\ }\bibfield  {title} {\enquote {\bibinfo {title}
  {{Stochastic backgrounds in alternative theories of gravity: overlap
  reduction functions for pulsar timing arrays}},}\ }\href {\doibase
  10.1103/PhysRevD.85.082001} {\bibfield  {journal} {\bibinfo  {journal} {Phys.
  Rev.}\ }\textbf {\bibinfo {volume} {D85}},\ \bibinfo {pages} {082001}
  (\bibinfo {year} {2012})},\ \Eprint {http://arxiv.org/abs/1111.5661}
  {arXiv:1111.5661 [astro-ph.HE]} \BibitemShut {NoStop}%
%%CITATION = ARXIV:1111.5661;%%
\bibitem [{\citenamefont {Gair}\ \emph {et~al.}(2014)\citenamefont {Gair},
  \citenamefont {Romano}, \citenamefont {Taylor},\ and\ \citenamefont
  {Mingarelli}}]{Gair:2014rwa}%
  \BibitemOpen
  \bibfield  {author} {\bibinfo {author} {\bibfnamefont {Jonathan}\
  \bibnamefont {Gair}}, \bibinfo {author} {\bibfnamefont {Joseph~D.}\
  \bibnamefont {Romano}}, \bibinfo {author} {\bibfnamefont {Stephen}\
  \bibnamefont {Taylor}}, \ and\ \bibinfo {author} {\bibfnamefont {Chiara
  M.~F.}\ \bibnamefont {Mingarelli}},\ }\bibfield  {title} {\enquote {\bibinfo
  {title} {{Mapping gravitational-wave backgrounds using methods from CMB
  analysis: Application to pulsar timing arrays}},}\ }\href {\doibase
  10.1103/PhysRevD.90.082001} {\bibfield  {journal} {\bibinfo  {journal} {Phys.
  Rev. D}\ }\textbf {\bibinfo {volume} {90}},\ \bibinfo {pages} {082001}
  (\bibinfo {year} {2014})},\ \Eprint {http://arxiv.org/abs/1406.4664}
  {arXiv:1406.4664 [gr-qc]} \BibitemShut {NoStop}%
\bibitem [{\citenamefont {Gair}\ \emph {et~al.}(2015)\citenamefont {Gair},
  \citenamefont {Romano},\ and\ \citenamefont {Taylor}}]{Gair:2015hra}%
  \BibitemOpen
  \bibfield  {author} {\bibinfo {author} {\bibfnamefont {Jonathan~R.}\
  \bibnamefont {Gair}}, \bibinfo {author} {\bibfnamefont {Joseph~D.}\
  \bibnamefont {Romano}}, \ and\ \bibinfo {author} {\bibfnamefont {Stephen~R.}\
  \bibnamefont {Taylor}},\ }\bibfield  {title} {\enquote {\bibinfo {title}
  {{Mapping gravitational-wave backgrounds of arbitrary polarisation using
  pulsar timing arrays}},}\ }\href {\doibase 10.1103/PhysRevD.92.102003}
  {\bibfield  {journal} {\bibinfo  {journal} {Phys. Rev. D}\ }\textbf {\bibinfo
  {volume} {92}},\ \bibinfo {pages} {102003} (\bibinfo {year} {2015})},\
  \Eprint {http://arxiv.org/abs/1506.08668} {arXiv:1506.08668 [gr-qc]}
  \BibitemShut {NoStop}%
\bibitem [{\citenamefont {Qin}\ \emph {et~al.}(2019)\citenamefont {Qin},
  \citenamefont {Boddy}, \citenamefont {Kamionkowski},\ and\ \citenamefont
  {Dai}}]{Qin:2018yhy}%
  \BibitemOpen
  \bibfield  {author} {\bibinfo {author} {\bibfnamefont {Wenzer}\ \bibnamefont
  {Qin}}, \bibinfo {author} {\bibfnamefont {Kimberly~K.}\ \bibnamefont
  {Boddy}}, \bibinfo {author} {\bibfnamefont {Marc}\ \bibnamefont
  {Kamionkowski}}, \ and\ \bibinfo {author} {\bibfnamefont {Liang}\
  \bibnamefont {Dai}},\ }\bibfield  {title} {\enquote {\bibinfo {title}
  {{Pulsar-timing arrays, astrometry, and gravitational waves}},}\ }\href
  {\doibase 10.1103/PhysRevD.99.063002} {\bibfield  {journal} {\bibinfo
  {journal} {Phys. Rev.}\ }\textbf {\bibinfo {volume} {D99}},\ \bibinfo {pages}
  {063002} (\bibinfo {year} {2019})},\ \Eprint
  {http://arxiv.org/abs/1810.02369} {arXiv:1810.02369 [astro-ph.CO]}
  \BibitemShut {NoStop}%
%%CITATION = ARXIV:1810.02369;%%
\bibitem [{\citenamefont {Mihaylov}\ \emph {et~al.}(2018)\citenamefont
  {Mihaylov}, \citenamefont {Moore}, \citenamefont {Gair}, \citenamefont
  {Lasenby},\ and\ \citenamefont {Gilmore}}]{Mihaylov:2018uqm}%
  \BibitemOpen
  \bibfield  {author} {\bibinfo {author} {\bibfnamefont {Deyan~P.}\
  \bibnamefont {Mihaylov}}, \bibinfo {author} {\bibfnamefont {Christopher~J.}\
  \bibnamefont {Moore}}, \bibinfo {author} {\bibfnamefont {Jonathan~R.}\
  \bibnamefont {Gair}}, \bibinfo {author} {\bibfnamefont {Anthony}\
  \bibnamefont {Lasenby}}, \ and\ \bibinfo {author} {\bibfnamefont {Gerard}\
  \bibnamefont {Gilmore}},\ }\bibfield  {title} {\enquote {\bibinfo {title}
  {{Astrometric Effects of Gravitational Wave Backgrounds with non-Einsteinian
  Polarizations}},}\ }\href {\doibase 10.1103/PhysRevD.97.124058} {\bibfield
  {journal} {\bibinfo  {journal} {Phys. Rev.}\ }\textbf {\bibinfo {volume}
  {D97}},\ \bibinfo {pages} {124058} (\bibinfo {year} {2018})},\ \Eprint
  {http://arxiv.org/abs/1804.00660} {arXiv:1804.00660 [gr-qc]} \BibitemShut
  {NoStop}%
%%CITATION = ARXIV:1804.00660;%%
\bibitem [{\citenamefont {O'Beirne}\ and\ \citenamefont
  {Cornish}(2018)}]{OBeirne:2018slh}%
  \BibitemOpen
  \bibfield  {author} {\bibinfo {author} {\bibfnamefont {Logan}\ \bibnamefont
  {O'Beirne}}\ and\ \bibinfo {author} {\bibfnamefont {Neil~J.}\ \bibnamefont
  {Cornish}},\ }\bibfield  {title} {\enquote {\bibinfo {title} {{Constraining
  the Polarization Content of Gravitational Waves with Astrometry}},}\ }\href
  {\doibase 10.1103/PhysRevD.98.024020} {\bibfield  {journal} {\bibinfo
  {journal} {Phys. Rev.}\ }\textbf {\bibinfo {volume} {D98}},\ \bibinfo {pages}
  {024020} (\bibinfo {year} {2018})},\ \Eprint
  {http://arxiv.org/abs/1804.03146} {arXiv:1804.03146 [gr-qc]} \BibitemShut
  {NoStop}%
%%CITATION = ARXIV:1804.03146;%%
\bibitem [{\citenamefont {Mihaylov}\ \emph {et~al.}(2019)\citenamefont
  {Mihaylov}, \citenamefont {Moore}, \citenamefont {Gair}, \citenamefont
  {Lasenby},\ and\ \citenamefont {Gilmore}}]{Mihaylov:2019lft}%
  \BibitemOpen
  \bibfield  {author} {\bibinfo {author} {\bibfnamefont {Deyan~P.}\
  \bibnamefont {Mihaylov}}, \bibinfo {author} {\bibfnamefont {Christopher~J.}\
  \bibnamefont {Moore}}, \bibinfo {author} {\bibfnamefont {Jonathan}\
  \bibnamefont {Gair}}, \bibinfo {author} {\bibfnamefont {Anthony}\
  \bibnamefont {Lasenby}}, \ and\ \bibinfo {author} {\bibfnamefont {Gerard}\
  \bibnamefont {Gilmore}},\ }\bibfield  {title} {\enquote {\bibinfo {title}
  {{Astrometric Effects of Gravitational Wave Backgrounds with non-Luminal
  Propagation Speeds}},}\ }\href@noop {} {\  (\bibinfo {year} {2019})},\
  \Eprint {http://arxiv.org/abs/1911.10356} {arXiv:1911.10356 [gr-qc]}
  \BibitemShut {NoStop}%
%%CITATION = ARXIV:1911.10356;%%
\bibitem [{\citenamefont {Cornish}\ \emph {et~al.}(2018)\citenamefont
  {Cornish}, \citenamefont {O'Beirne}, \citenamefont {Taylor},\ and\
  \citenamefont {Yunes}}]{Cornish:2017oic}%
  \BibitemOpen
  \bibfield  {author} {\bibinfo {author} {\bibfnamefont {Neil~J.}\ \bibnamefont
  {Cornish}}, \bibinfo {author} {\bibfnamefont {Logan}\ \bibnamefont
  {O'Beirne}}, \bibinfo {author} {\bibfnamefont {Stephen~R.}\ \bibnamefont
  {Taylor}}, \ and\ \bibinfo {author} {\bibfnamefont {Nicol\'as}\ \bibnamefont
  {Yunes}},\ }\bibfield  {title} {\enquote {\bibinfo {title} {{Constraining
  alternative theories of gravity using pulsar timing arrays}},}\ }\href
  {\doibase 10.1103/PhysRevLett.120.181101} {\bibfield  {journal} {\bibinfo
  {journal} {Phys. Rev. Lett.}\ }\textbf {\bibinfo {volume} {120}},\ \bibinfo
  {pages} {181101} (\bibinfo {year} {2018})},\ \Eprint
  {http://arxiv.org/abs/1712.07132} {arXiv:1712.07132 [gr-qc]} \BibitemShut
  {NoStop}%
\bibitem [{\citenamefont {O'Beirne}\ \emph {et~al.}(2019)\citenamefont
  {O'Beirne}, \citenamefont {Cornish}, \citenamefont {Vigeland},\ and\
  \citenamefont {Taylor}}]{OBeirne:2019lwp}%
  \BibitemOpen
  \bibfield  {author} {\bibinfo {author} {\bibfnamefont {Logan}\ \bibnamefont
  {O'Beirne}}, \bibinfo {author} {\bibfnamefont {Neil~J.}\ \bibnamefont
  {Cornish}}, \bibinfo {author} {\bibfnamefont {Sarah~J.}\ \bibnamefont
  {Vigeland}}, \ and\ \bibinfo {author} {\bibfnamefont {Stephen~R.}\
  \bibnamefont {Taylor}},\ }\bibfield  {title} {\enquote {\bibinfo {title}
  {{Constraining alternative polarization states of gravitational waves from
  individual black hole binaries using pulsar timing arrays}},}\ }\href
  {\doibase 10.1103/PhysRevD.99.124039} {\bibfield  {journal} {\bibinfo
  {journal} {Phys. Rev. D}\ }\textbf {\bibinfo {volume} {99}},\ \bibinfo
  {pages} {124039} (\bibinfo {year} {2019})},\ \Eprint
  {http://arxiv.org/abs/1904.02744} {arXiv:1904.02744 [gr-qc]} \BibitemShut
  {NoStop}%
\bibitem [{\citenamefont {Baskaran}\ \emph {et~al.}(2008)\citenamefont
  {Baskaran}, \citenamefont {Polnarev}, \citenamefont {Pshirkov},\ and\
  \citenamefont {Postnov}}]{Baskaran:2008za}%
  \BibitemOpen
  \bibfield  {author} {\bibinfo {author} {\bibfnamefont {D.}~\bibnamefont
  {Baskaran}}, \bibinfo {author} {\bibfnamefont {A.G.}\ \bibnamefont
  {Polnarev}}, \bibinfo {author} {\bibfnamefont {M.S.}\ \bibnamefont
  {Pshirkov}}, \ and\ \bibinfo {author} {\bibfnamefont {K.A.}\ \bibnamefont
  {Postnov}},\ }\bibfield  {title} {\enquote {\bibinfo {title} {{Limits on the
  speed of gravitational waves from pulsar timing}},}\ }\href {\doibase
  10.1103/PhysRevD.78.044018} {\bibfield  {journal} {\bibinfo  {journal} {Phys.
  Rev. D}\ }\textbf {\bibinfo {volume} {78}},\ \bibinfo {pages} {044018}
  (\bibinfo {year} {2008})},\ \Eprint {http://arxiv.org/abs/0805.3103}
  {arXiv:0805.3103 [astro-ph]} \BibitemShut {NoStop}%
\bibitem [{\citenamefont {Lee}\ \emph {et~al.}(2010)\citenamefont {Lee},
  \citenamefont {Jenet}, \citenamefont {Price}, \citenamefont {Wex},\ and\
  \citenamefont {Kramer}}]{Lee:2010cg}%
  \BibitemOpen
  \bibfield  {author} {\bibinfo {author} {\bibfnamefont {Kejia}\ \bibnamefont
  {Lee}}, \bibinfo {author} {\bibfnamefont {Fredrick~A.}\ \bibnamefont
  {Jenet}}, \bibinfo {author} {\bibfnamefont {Richard~H.}\ \bibnamefont
  {Price}}, \bibinfo {author} {\bibfnamefont {Norbert}\ \bibnamefont {Wex}}, \
  and\ \bibinfo {author} {\bibfnamefont {Michael}\ \bibnamefont {Kramer}},\
  }\bibfield  {title} {\enquote {\bibinfo {title} {{Detecting massive gravitons
  using pulsar timing arrays}},}\ }\href {\doibase
  10.1088/0004-637X/722/2/1589} {\bibfield  {journal} {\bibinfo  {journal}
  {Astrophys. J.}\ }\textbf {\bibinfo {volume} {722}},\ \bibinfo {pages}
  {1589--1597} (\bibinfo {year} {2010})},\ \Eprint
  {http://arxiv.org/abs/1008.2561} {arXiv:1008.2561 [astro-ph.HE]} \BibitemShut
  {NoStop}%
\bibitem [{\citenamefont {Lee}(2013)}]{Lee:2014awa}%
  \BibitemOpen
  \bibfield  {author} {\bibinfo {author} {\bibfnamefont {K.J.}\ \bibnamefont
  {Lee}},\ }\bibfield  {title} {\enquote {\bibinfo {title} {{Pulsar Timing
  Arrays and Gravity Tests in the Radiative Regime}},}\ }\href {\doibase
  10.1088/0264-9381/30/22/224016} {\bibfield  {journal} {\bibinfo  {journal}
  {Class. Quant. Grav.}\ }\textbf {\bibinfo {volume} {30}},\ \bibinfo {pages}
  {224016} (\bibinfo {year} {2013})},\ \Eprint {http://arxiv.org/abs/1404.2090}
  {arXiv:1404.2090 [astro-ph.CO]} \BibitemShut {NoStop}%
\bibitem [{\citenamefont {Dai}\ \emph {et~al.}(2012)\citenamefont {Dai},
  \citenamefont {Kamionkowski},\ and\ \citenamefont {Jeong}}]{Dai:2012bc}%
  \BibitemOpen
  \bibfield  {author} {\bibinfo {author} {\bibfnamefont {Liang}\ \bibnamefont
  {Dai}}, \bibinfo {author} {\bibfnamefont {Marc}\ \bibnamefont
  {Kamionkowski}}, \ and\ \bibinfo {author} {\bibfnamefont {Donghui}\
  \bibnamefont {Jeong}},\ }\bibfield  {title} {\enquote {\bibinfo {title}
  {{Total Angular Momentum Waves for Scalar, Vector, and Tensor Fields}},}\
  }\href {\doibase 10.1103/PhysRevD.86.125013} {\bibfield  {journal} {\bibinfo
  {journal} {Phys. Rev.}\ }\textbf {\bibinfo {volume} {D86}},\ \bibinfo {pages}
  {125013} (\bibinfo {year} {2012})},\ \Eprint {http://arxiv.org/abs/1209.0761}
  {arXiv:1209.0761 [astro-ph.CO]} \BibitemShut {NoStop}%
%%CITATION = ARXIV:1209.0761;%%
\bibitem [{\citenamefont {Abbott}\ \emph
  {et~al.}(2017{\natexlab{a}})\citenamefont {Abbott} \emph
  {et~al.}}]{Monitor:2017mdv}%
  \BibitemOpen
  \bibfield  {author} {\bibinfo {author} {\bibfnamefont {B.~P.}\ \bibnamefont
  {Abbott}} \emph {et~al.} (\bibinfo {collaboration} {LIGO Scientific, Virgo,
  Fermi-GBM, INTEGRAL}),\ }\bibfield  {title} {\enquote {\bibinfo {title}
  {{Gravitational Waves and Gamma-rays from a Binary Neutron Star Merger:
  GW170817 and GRB 170817A}},}\ }\href {\doibase 10.3847/2041-8213/aa920c}
  {\bibfield  {journal} {\bibinfo  {journal} {Astrophys. J.}\ }\textbf
  {\bibinfo {volume} {848}},\ \bibinfo {pages} {L13} (\bibinfo {year}
  {2017}{\natexlab{a}})},\ \Eprint {http://arxiv.org/abs/1710.05834}
  {arXiv:1710.05834 [astro-ph.HE]} \BibitemShut {NoStop}%
%%CITATION = ARXIV:1710.05834;%%
\bibitem [{\citenamefont {Abbott}\ \emph
  {et~al.}(2017{\natexlab{b}})\citenamefont {Abbott} \emph
  {et~al.}}]{Abbott:2017oio}%
  \BibitemOpen
  \bibfield  {author} {\bibinfo {author} {\bibfnamefont {B.P.}\ \bibnamefont
  {Abbott}} \emph {et~al.} (\bibinfo {collaboration} {LIGO Scientific,
  Virgo}),\ }\bibfield  {title} {\enquote {\bibinfo {title} {{GW170814: A
  Three-Detector Observation of Gravitational Waves from a Binary Black Hole
  Coalescence}},}\ }\href {\doibase 10.1103/PhysRevLett.119.141101} {\bibfield
  {journal} {\bibinfo  {journal} {Phys. Rev. Lett.}\ }\textbf {\bibinfo
  {volume} {119}},\ \bibinfo {pages} {141101} (\bibinfo {year}
  {2017}{\natexlab{b}})},\ \Eprint {http://arxiv.org/abs/1709.09660}
  {arXiv:1709.09660 [gr-qc]} \BibitemShut {NoStop}%
\bibitem [{\citenamefont {Isi}\ and\ \citenamefont
  {Weinstein}(2017)}]{Isi:2017fbj}%
  \BibitemOpen
  \bibfield  {author} {\bibinfo {author} {\bibfnamefont {Maximiliano}\
  \bibnamefont {Isi}}\ and\ \bibinfo {author} {\bibfnamefont {Alan~J.}\
  \bibnamefont {Weinstein}},\ }\bibfield  {title} {\enquote {\bibinfo {title}
  {{Probing gravitational wave polarizations with signals from compact binary
  coalescences}},}\ }\href@noop {} {\  (\bibinfo {year} {2017})},\ \Eprint
  {http://arxiv.org/abs/1710.03794} {arXiv:1710.03794 [gr-qc]} \BibitemShut
  {NoStop}%
\bibitem [{\citenamefont {Abbott}\ \emph {et~al.}(2018)\citenamefont {Abbott}
  \emph {et~al.}}]{Abbott:2018utx}%
  \BibitemOpen
  \bibfield  {author} {\bibinfo {author} {\bibfnamefont {Benjamin~P.}\
  \bibnamefont {Abbott}} \emph {et~al.} (\bibinfo {collaboration} {LIGO
  Scientific, Virgo}),\ }\bibfield  {title} {\enquote {\bibinfo {title}
  {{Search for Tensor, Vector, and Scalar Polarizations in the Stochastic
  Gravitational-Wave Background}},}\ }\href {\doibase
  10.1103/PhysRevLett.120.201102} {\bibfield  {journal} {\bibinfo  {journal}
  {Phys. Rev. Lett.}\ }\textbf {\bibinfo {volume} {120}},\ \bibinfo {pages}
  {201102} (\bibinfo {year} {2018})},\ \Eprint
  {http://arxiv.org/abs/1802.10194} {arXiv:1802.10194 [gr-qc]} \BibitemShut
  {NoStop}%
\bibitem [{\citenamefont {Arzoumanian}\ \emph {et~al.}(2018)\citenamefont
  {Arzoumanian} \emph {et~al.}}]{Arzoumanian:2018saf}%
  \BibitemOpen
  \bibfield  {author} {\bibinfo {author} {\bibfnamefont {Z.}~\bibnamefont
  {Arzoumanian}} \emph {et~al.} (\bibinfo {collaboration} {NANOGRAV}),\
  }\bibfield  {title} {\enquote {\bibinfo {title} {{The NANOGrav 11-year Data
  Set: Pulsar-timing Constraints On The Stochastic Gravitational-wave
  Background}},}\ }\href {\doibase 10.3847/1538-4357/aabd3b} {\bibfield
  {journal} {\bibinfo  {journal} {Astrophys. J.}\ }\textbf {\bibinfo {volume}
  {859}},\ \bibinfo {pages} {47} (\bibinfo {year} {2018})},\ \Eprint
  {http://arxiv.org/abs/1801.02617} {arXiv:1801.02617 [astro-ph.HE]}
  \BibitemShut {NoStop}%
%%CITATION = ARXIV:1801.02617;%%
\bibitem [{\citenamefont {Mingarelli}\ and\ \citenamefont
  {Sidery}(2014)}]{Mingarelli:2014xfa}%
  \BibitemOpen
  \bibfield  {author} {\bibinfo {author} {\bibfnamefont {Chiara M.~F.}\
  \bibnamefont {Mingarelli}}\ and\ \bibinfo {author} {\bibfnamefont {Trevor}\
  \bibnamefont {Sidery}},\ }\bibfield  {title} {\enquote {\bibinfo {title}
  {{Effect of small interpulsar distances in stochastic gravitational wave
  background searches with pulsar timing arrays}},}\ }\href {\doibase
  10.1103/PhysRevD.90.062011} {\bibfield  {journal} {\bibinfo  {journal} {Phys.
  Rev.}\ }\textbf {\bibinfo {volume} {D90}},\ \bibinfo {pages} {062011}
  (\bibinfo {year} {2014})},\ \Eprint {http://arxiv.org/abs/1408.6840}
  {arXiv:1408.6840 [astro-ph.HE]} \BibitemShut {NoStop}%
%%CITATION = ARXIV:1408.6840;%%
\bibitem [{\citenamefont {Chiba}\ \emph {et~al.}(2007)\citenamefont {Chiba},
  \citenamefont {Smith},\ and\ \citenamefont {Erickcek}}]{Chiba:2006jp}%
  \BibitemOpen
  \bibfield  {author} {\bibinfo {author} {\bibfnamefont {Takeshi}\ \bibnamefont
  {Chiba}}, \bibinfo {author} {\bibfnamefont {Tristan~L.}\ \bibnamefont
  {Smith}}, \ and\ \bibinfo {author} {\bibfnamefont {Adrienne~L.}\ \bibnamefont
  {Erickcek}},\ }\bibfield  {title} {\enquote {\bibinfo {title} {{Solar System
  constraints to general f(R) gravity}},}\ }\href {\doibase
  10.1103/PhysRevD.75.124014} {\bibfield  {journal} {\bibinfo  {journal} {Phys.
  Rev.}\ }\textbf {\bibinfo {volume} {D75}},\ \bibinfo {pages} {124014}
  (\bibinfo {year} {2007})},\ \Eprint {http://arxiv.org/abs/astro-ph/0611867}
  {arXiv:astro-ph/0611867 [astro-ph]} \BibitemShut {NoStop}%
%%CITATION = ASTRO-PH/0611867;%%
\bibitem [{\citenamefont {O'Hanlon}(1972)}]{OHanlon:1972xqa}%
  \BibitemOpen
  \bibfield  {author} {\bibinfo {author} {\bibfnamefont {John}\ \bibnamefont
  {O'Hanlon}},\ }\bibfield  {title} {\enquote {\bibinfo {title}
  {{Intermediate-range gravity - a generally covariant model}},}\ }\href
  {\doibase 10.1103/PhysRevLett.29.137} {\bibfield  {journal} {\bibinfo
  {journal} {Phys.\ Rev.\ Lett.}\ }\textbf {\bibinfo {volume} {29}},\ \bibinfo
  {pages} {137--138} (\bibinfo {year} {1972})}\BibitemShut {NoStop}%
\bibitem [{\citenamefont {Teyssandier}\ and\ \citenamefont
  {Tourrenc}(1983)}]{Teyssandier:1983zz}%
  \BibitemOpen
  \bibfield  {author} {\bibinfo {author} {\bibfnamefont {P.}~\bibnamefont
  {Teyssandier}}\ and\ \bibinfo {author} {\bibfnamefont {Ph.}\ \bibnamefont
  {Tourrenc}},\ }\bibfield  {title} {\enquote {\bibinfo {title} {{The Cauchy
  problem for the R+R**2 theories of gravity without torsion}},}\ }\href
  {\doibase 10.1063/1.525659} {\bibfield  {journal} {\bibinfo  {journal} {J.\
  Math.\ Phys.}\ }\textbf {\bibinfo {volume} {24}},\ \bibinfo {pages} {2793}
  (\bibinfo {year} {1983})}\BibitemShut {NoStop}%
\bibitem [{\citenamefont {Liang}\ \emph {et~al.}(2017)\citenamefont {Liang},
  \citenamefont {Gong}, \citenamefont {Hou},\ and\ \citenamefont
  {Liu}}]{Liang:2017ahj}%
  \BibitemOpen
  \bibfield  {author} {\bibinfo {author} {\bibfnamefont {Dicong}\ \bibnamefont
  {Liang}}, \bibinfo {author} {\bibfnamefont {Yungui}\ \bibnamefont {Gong}},
  \bibinfo {author} {\bibfnamefont {Shaoqi}\ \bibnamefont {Hou}}, \ and\
  \bibinfo {author} {\bibfnamefont {Yunqi}\ \bibnamefont {Liu}},\ }\bibfield
  {title} {\enquote {\bibinfo {title} {{Polarizations of gravitational waves in
  $f(R)$ gravity}},}\ }\href {\doibase 10.1103/PhysRevD.95.104034} {\bibfield
  {journal} {\bibinfo  {journal} {Phys. Rev.}\ }\textbf {\bibinfo {volume}
  {D95}},\ \bibinfo {pages} {104034} (\bibinfo {year} {2017})},\ \Eprint
  {http://arxiv.org/abs/1701.05998} {arXiv:1701.05998 [gr-qc]} \BibitemShut
  {NoStop}%
%%CITATION = ARXIV:1701.05998;%%
\bibitem [{\citenamefont {Moretti}\ \emph {et~al.}(2019)\citenamefont
  {Moretti}, \citenamefont {Bombacigno},\ and\ \citenamefont
  {Montani}}]{Moretti:2019yhs}%
  \BibitemOpen
  \bibfield  {author} {\bibinfo {author} {\bibfnamefont {Fabio}\ \bibnamefont
  {Moretti}}, \bibinfo {author} {\bibfnamefont {Flavio}\ \bibnamefont
  {Bombacigno}}, \ and\ \bibinfo {author} {\bibfnamefont {Giovanni}\
  \bibnamefont {Montani}},\ }\bibfield  {title} {\enquote {\bibinfo {title}
  {{Gauge invariant formulation of metric $f(R)$ gravity for gravitational
  waves}},}\ }\href {\doibase 10.1103/PhysRevD.100.084014} {\bibfield
  {journal} {\bibinfo  {journal} {Phys. Rev. D}\ }\textbf {\bibinfo {volume}
  {100}},\ \bibinfo {pages} {084014} (\bibinfo {year} {2019})},\ \Eprint
  {http://arxiv.org/abs/1906.01899} {arXiv:1906.01899 [gr-qc]} \BibitemShut
  {NoStop}%
\bibitem [{\citenamefont {Isi}\ and\ \citenamefont
  {Stein}(2018)}]{Isi:2018miq}%
  \BibitemOpen
  \bibfield  {author} {\bibinfo {author} {\bibfnamefont {Maximiliano}\
  \bibnamefont {Isi}}\ and\ \bibinfo {author} {\bibfnamefont {Leo~C.}\
  \bibnamefont {Stein}},\ }\bibfield  {title} {\enquote {\bibinfo {title}
  {{Measuring stochastic gravitational-wave energy beyond general
  relativity}},}\ }\href {\doibase 10.1103/PhysRevD.98.104025} {\bibfield
  {journal} {\bibinfo  {journal} {Phys. Rev. D}\ }\textbf {\bibinfo {volume}
  {98}},\ \bibinfo {pages} {104025} (\bibinfo {year} {2018})},\ \Eprint
  {http://arxiv.org/abs/1807.02123} {arXiv:1807.02123 [gr-qc]} \BibitemShut
  {NoStop}%
\end{thebibliography}%

\end{document}